\documentclass[preprint,3p,12pt]{elsarticle}
\usepackage{mathrsfs}
\usepackage{amsmath}
\usepackage{stmaryrd}
\usepackage{bbding}
\usepackage{dcolumn}
\usepackage{graphicx}
\usepackage{amsfonts}
\usepackage{amssymb}
\usepackage{psfrag}
\usepackage{wrapfig}
\usepackage{subfigure}
\usepackage{makeidx}
\usepackage{bm}
\usepackage{epsf}
\usepackage{color}
\usepackage{epsfig}
\usepackage{setspace}
\usepackage{graphicx}
\usepackage{epstopdf}
\usepackage{psfrag}
\usepackage{subfigure}
\usepackage{multirow}
\usepackage{diagbox}
\usepackage{verbatim}

\usepackage{makecell}
\usepackage{float}
\usepackage{esint}
\usepackage{comment}
\usepackage{movie15}
\usepackage{hyperref}
\usepackage[ruled,linesnumbered]{algorithm2e}
\usepackage{algorithmicx}
\usepackage[noend]{algpseudocode}

\newcommand{\mathsym}[1]{{}}
\newcommand{\unicode}[1]{{}}
\begin{document}
	
\title{Implicit unified gas kinetic particle method for steady-state solution of multiscale phonon transport}

	\author[HKUST1]{Hongyu Liu}
    \ead{hliudv@connect.ust.hk}
	
        \author[HKUST1]{Xiaojian Yang}
	\ead{xyangbm@connect.ust.hk}

    \author[add1]{Chuang Zhang}
\ead{zhangc520@hdu.edu.cn}

    	\author[XJTU]{Xing Ji}
	\ead{jixing@xjtu.edu.cn}
	
	\author[HKUST1,HKUST2,HKUST3]{Kun Xu\corref{cor1}}
	\ead{makxu@ust.hk}
	
	\address[HKUST1]{Department of Mathematics, Hong Kong University of Science and Technology, Clear Water Bay, Kowloon, Hong Kong}

    	\address[HKUST2]{Department of Mechanical and Aerospace Engineering, Hong Kong University of Science and Technology, Clear Water Bay, Kowloon, Hong Kong}

        	\address[HKUST3]{Shenzhen Research Institute, Hong Kong University of Science and Technology, Shenzhen, China}
	\cortext[cor1]{Corresponding author}

	\address[XJTU]{Shaanxi Key Laboratory of Environment and Control for Flight Vehicle, Xi'an Jiaotong University, Xi'an, China}

    \address[add1]{Institute of Energy, School of Sciences, Hangzhou Dianzi University, Hangzhou 310018, China}

\begin{abstract}

This paper presents a highly efficient implicit unified gas-kinetic particle (IUGKP) method for obtaining steady-state solutions of multi-scale phonon transport. The method adapts and reinterprets the integral solution of the BGK equation for time-independent solutions. The distribution function at a given point is determined solely by the surrounding equilibrium states, where the corresponding macroscopic quantities are computed through a weighted sum of equilibrium distribution functions from neighboring spatial positions.
From a particle perspective, changes in macroscopic quantities within a cell result from particle transport across cell interfaces. These particles are sampled according to the equilibrium state of their original cells, accounting for their mean free path as the traveling distance. The IUGKP method evolves the solution according to the physical relaxation time scale, achieving high efficiency in large Knudsen number regimes.
To accelerate convergence for small Knudsen numbers, an inexact Newton iteration method is implemented, incorporating macroscopic equations for convergence acceleration in the near-diffusive limit. The method also addresses spatial-temporal inconsistency caused by relaxation time variations in physical space through the null-collision concept.
Numerical tests demonstrate the method's excellent performance in accelerating multi-scale phonon transport solutions, achieving speedups of one to two orders of magnitude. The IUGKP method proves to be an efficient and accurate computational tool for simulating multiscale non-equilibrium heat transfer, offering significant advantages over traditional methods in both numerical performance and physical applicability.

\end{abstract}


\maketitle

\section{Introduction}
For heat transfer phenomena at the micro- and nanoscales, Fourier's law of heat conduction is not applicable.
It fails to capture non-equilibrium
thermal conduction phenomena such as ballistic transport, nonlocal, nonlinear, size effects, and complex interfacial scattering mechanisms~\cite{phononsnanoscale, RevModPhys.90.041002, RevModPhys.94.025002}.
Therefore, developing heat conduction models that capture phonon drift, scattering, absorption, and emission is crucial. The phonon Boltzmann transport equation (BTE) \cite{murthy2005review, ziman2001electrons, chattopadhyay2014comparative} offers a framework that overcomes the limitation of Fourier’s law and accurately reflects the complex transport mechanisms in modern electronic devices.
However, analytical solutions of the BTE for realistic electronic devices are extremely challenging due to the multiscale nature of phonon transport and complex interactions with materials.
Consequently, numerical simulations become indispensable, necessitating the development of advanced computational methods to accurately and efficiently model phonon transport for improved thermal management in electronic devices.
The BTE has two angular dimensions, one frequency dimension, and three spatial dimensions~\cite{barry2022boltzmann, mazumder_boltzmann_2022}.
These six dimensions result in enormous computational cost, posing significant challenges for numerical simulations~\cite{guo_progress_DUGKS, peraud_monte_2014, lacroix2005monte}.

Over the past few decades, two main categories of methods have been developed for solving the phonon BTE: deterministic methods, represented by the DOM~\cite{stamnes1988dom, murthy1998domunstruct, SyedAA14LargeScale, FivelandVA96Acceleration}, and stochastic particle methods, represented by the MC method~\cite{DSMC_book_1994, DSMC_phonon_1994, mazumder2001monte, lacroix2005monte,mittal2010monte, PJP11MC, peraud2015derivationmonte, peraud_monte_2014}.
The DOM method decouples particle free transport and collision effects in numerical handling, which requires the time step to be smaller than the particle relaxation time.
It exhibits significant numerical dissipation at low Knudsen numbers and requires a large number of computational steps to reach a converged solution.
Traditional explicit Monte Carlo methods significantly reduce the number of velocity space elements.
However, they also decouple particle transport from collisions. Similar to the DOM method, a large number of computational steps are required to achieve convergence at low Knudsen numbers~\cite{peraud_monte_2014, lacroix2005monte, PJP11MC, SILVA2024108954, PATHAK2021108003}.
To overcome the limitations of the aforementioned methods at low Knudsen numbers, deterministic methods with multiscale properties, exemplified by unified gas-kinetic scheme (UGKS) and discrete UGKS (DUGKS)~\cite{UGKS,guo_progress_DUGKS, guo2016dugksphonon, luo2017dugksphonon, zhang2019dugksphonon}, have been developed in recent years and successfully applied to phonon transport.
At the same time, the general synthetic iterative scheme (GSIS) \cite{Chuang17gray, zhang2023acceleration, zhang2025synthetic} has also achieved significant success in solving phonon transport problems.
For the multiscale particle method, the unified gas-kinetic particle (UGKP) and unified gas-kinetic wave-particle (UGKWP) methods have been widely applied to multiscale neutral gas transport \cite{zhu2019ugkwp}, plasmas \cite{liu2021ugkwp, pu2025ugkwp}, multiphase flows \cite{yang2024solid}, radiation \cite{yang2025rad}, turbulence simulation \cite{yang2025turb} and phonon transport \cite{liu2025unified}.

Deterministic multiscale methods have also developed corresponding steady-state acceleration techniques, such as implicit UGKS and implicit DUGKS \cite{zhu2016implicit, zhang2023acceleration}. Compared to explicit schemes, implicit schemes significantly enhance the convergence rate; however, because they still require discretization of a large velocity space, they ultimately consume a substantial amount of computational time.
Statistical multi-scale methods, such as UGKWP, are bounded by numerical time steps, and it's relatively hard to develop an implicit scheme for further accelerating their convergence speed.
Recently, implicit UGKWP~\cite{liu2023implicit, hu2024implicit} has been gradually applied in the field of radiation, which may offer some insights for accelerating other multiscale transport UGKWP methods.

This paper proposes a novel particle method to efficiently solve the phonon transport equation. The method combines the integral solution of the steady-state phonon BGK equation with a null collision approach and macroscopic prediction equations.
The fundamental principle is that the distribution function at any point is determined solely by equilibrium states at surrounding locations. At large Knudsen numbers, particles can travel distances significantly larger than those in numerical time-step-based transport, resulting in much faster convergence compared to conventional Monte Carlo (MC) methods.
For low Knudsen numbers, similar to the IUGKS method, macroscopic prediction equations determine the new equilibrium state for particles sampling in the next iteration. This approach compensates for tracking particles with short mean free paths under small Knudsen number condition.
Unlike existing steady-state Monte Carlo method for phonon transport, our approach resolves the spatio-temporal inconsistency caused by varying relaxation times in physical space through the null-collision concept.
The method maintains the high efficiency of steady-state Monte Carlo methods at large Knudsen numbers while demonstrating superior multiscale capability - a feature absent in current phonon Monte Carlo methods. Additionally, it enhances computational efficiency at low Knudsen numbers by using macroscopic equations to predict collision equilibrium states, which guide particle resampling in subsequent steps.

This paper is organized as follows. Section 2 introduces the steady phonon BGK equation and its integral solution. Section 3 presents the details of the IGKP method. Section 4 is the numerical examples. The last section is the conclusion.

\section{Steady phonon BGK Equation}
In general, the phonon BTE can be simplified using the BGK type relaxation time approximation model \cite{BGK, ziman2001electrons, Chuang17gray, PJP11MC}
\begin{equation}
\boldsymbol{V_g} \cdot \nabla f=\frac{1}{\tau}\left(f^{e q}-f\right),
\label{eq:BGKBTE}
\end{equation}
where $f$ is the phonon distribution function, $\boldsymbol{V_g}=|\boldsymbol{V_g}| \boldsymbol{s}$ is the group velocity, $\boldsymbol{s}= (\cos \theta, \\ \sin \theta \cos \varphi, \sin \theta \sin \varphi)$ is the unit directional vector $(\theta$ is the polar angle and $\varphi$ is the azimuthal angle), $\tau$ is the relaxation time, $f^{eq}$ is the equilibrium distribution function, which satisfy the Bose-Einstein distribution,
\begin{equation}\label{bose-einstein}
f^{eq} (T)=\frac{1}{\exp \left(\hbar \omega / k_B T\right)-1},
\end{equation}
where $\hbar$ is the reduced Planck constant, $\omega$ is the frequency, $k_B$ is the Boltzmann constant and $T$ is the temperature.

In this paper, we employ the gray model, which neglects phonon dispersions and polarization characteristics and introduces the assumption that the phonon group velocity and relaxation time are constants.
Although this simplification cannot accurately capture frequency-dependent or anisotropic phonon transport behaviors in solid materials, it provides insights into phonon transport phenomena, such as ballistic transport and boundary scattering effects.
The Knudsen number of the system is defined as $\mathrm{Kn}=\lambda / L_0$, where $L_0$ is the characteristic length of the system and $\lambda=\boldsymbol{|V_g|} \tau$ is the phonon mean free path.
Eq.~\eqref{eq:BGKBTE} can also be expressed in terms of the phonon energy density per unit solid angle
\begin{equation}
\boldsymbol{V_g} \cdot \nabla e=\frac{1}{\tau}\left(e^{e q}-e\right).
\end{equation}
where
\begin{equation} \label{derivation function}
e(\boldsymbol{x}, \boldsymbol{s}, \omega, p)=
\sum_p \int   \left( \hbar \omega\left( f-f^{e q}\left(T_{\mathrm{ref}}\right)\right) D(\omega, p) / 4 \pi \right)   \mathrm{~d} \omega,
\end{equation}
\begin{equation} \label{derivation equilibrium}
e^{e q}(\boldsymbol{x}, \boldsymbol{s}, \omega, p)=\sum_p \int   \left( \hbar \omega\left(  f^{eq} (T) -f^{e q}\left(T_{\mathrm{ref}}\right)  \right) D(\omega, p) / 4 \pi \right)   \mathrm{~d} \omega,
\end{equation}
where $D(\omega, p)$ is the phonon density of state and $T_{\mathrm{ref}}$ is the reference temperature.
Taking a first-order Taylor expansion of distribution function at $T_0$, Eq.~\eqref{derivation equilibrium} can be expressed as
\begin{equation}\label{derivation final equilibrium}
    e^{eq} \approx \frac{C  \left(T-T_{\mathrm{ref}}\right)}{4 \pi},
\end{equation}
where $C $ is the volumetric specific heat.
The local energy $E$, temperature $T$ and heat flux $\boldsymbol{q}$ are obtained by taking the moments of the phonon distribution function of the energy density over the whole solid angle space,
\begin{align}
E  &=  \iint_{4 \pi} e \mathrm{~d} \Omega  ,  \\
T  &= \frac{ E }{C}  +T_{\mathrm{ref}} ,  \\
\boldsymbol{q} &= \iint_{4 \pi} \boldsymbol{V_g} e \mathrm{~d} \Omega.
\end{align}
Then, the integral solution can be obtained:
\begin{equation}
e(x, u)=\int_{x_0}^x \frac{e^{eq}}{u \tau\left(x^{\prime}\right)} \mathrm{e}^{-\int_{x^{\prime}}^x \frac{1}{u \tau\left(x^{\prime \prime}\right)} \mathrm{d} x^{\prime \prime}} \mathrm{d} x^{\prime}+e\left(x_0, u\right) \mathrm{e}^{-\int_{x_0}^x \frac{1}{u \tau\left(x^{\prime}\right)} \mathrm{d} x^{\prime}},
\end{equation}
where $x_0$ is the starting point and the relaxation time $\tau$ is related to space location $x$, u is the magnitude of the group velocity.

The integral solution of the steady BGK equation demonstrates that the particles located at the starting point have a  cumulative density function (CDF) $P_f$ to maintain their distribution while free-streaming to the endpoint $x$; the formulation of $P_f\left(x_0,x\right)$ is:
\begin{equation}
    P_f\left(x_0,x\right)=\mathrm{e}^{-\int_{x_0}^x \frac{1}{u \tau\left(x^{\prime}\right)} \mathrm{d} x^{\prime}}.
\end{equation}
Conversely, during transport, the probability density function (PDF) of experiencing a collision at point $x\prime$ and keeping the local equilibrium distribution to $x$ is:
\begin{equation}
P_c\left(x^\prime,x\right)=\frac{1}{u \tau\left(x^{\prime}\right)} \mathrm{e}^{-\int_{x^{\prime}}^x \frac{1}{u \tau\left(x^{\prime \prime}\right)} \mathrm{d} x^{\prime \prime}}.
\end{equation}
Considering the composition of the distribution function at a certain point $x$, the final distribution function at a certain point $x$ can be regarded as the mathematical expectation of the distribution functions from the starting point $x_0$ to that point $x$, weighted by their respective probabilities.

In other words, the distribution function at that point $x$ is a convex combination of the equilibrium distribution functions from itself and other positions and the boundary distribution function.
Suppose the boundary distribution function is at equilibrium. In that case, the distribution function at that point is a convex combination of the equilibrium distribution functions from all the points in the computational region:
\begin{equation}\label{sum-g}
    e_{i, k}^{n+1} = \sum_{j=0}^{N}\omega_{j,k}e^{eq,n}_{j,k},
\end{equation}
where N is the total number of cells of the grid.
Considering the case where $\tau$ is constant, and based on previous results that yield $P_c=\frac{dP_f}{dx}$, we can further rewrite the above equation as:
\begin{equation}\label{sum-g-CDF}
    e_{i, k}^{n+1}\left(x,u\right) = \sum \mathbf F\left({P_f\left(x_j,x\right)}\right)e^{eq,n}_{j,k}\left(x_j,u\right),
\end{equation}
where the summation takes over all particles with the specified velocity capable of traveling from point $x_j$ to point $x$ and $\mathbf{F}$ is the function related to the CDF.
This indicates that the distribution function at point x is obtained by probabilistically accumulating the corresponding equilibrium distribution functions from other points using the CDF.
\section{Steady-state acceleration in IUGKP}
Based on the previous introduction, the remainder of this paper will explain, from a Monte Carlo perspective, how particles can represent the transport and collision phenomena inherent in the integral solution.
\subsection{Constant $\tau$ case}
For convenience, we will initially assume that $\tau$ is constant; a method to handle variable $\tau$ will be presented later.
From the emission perspective: particles emitted from a given point in local equilibrium have free transport distances that follow the cumulative PDF $P_f$.
From the inverse sampling method, its free transport length is:
\begin{equation}
    \lambda = -u\tau ln\left(\eta\right).
\end{equation}
After traveling this distance, the particle undergoes a collision at the local point and then conforms to the local equilibrium distribution.
Then, Eq.~\eqref {sum-g-CDF} can be reformulated into a particle-based representation as follows:
\begin{equation}\label{sum-g-particle}
    W_{i} = \sum w^p.
\end{equation}
The physical picture is shown in Fig.~\ref{constant-tau-integral-particle}, illustrating the above process.
\begin{figure}[htp]	\label{constant-tau-integral-particle}
	\centering	
    \includegraphics[height=0.45\textwidth]{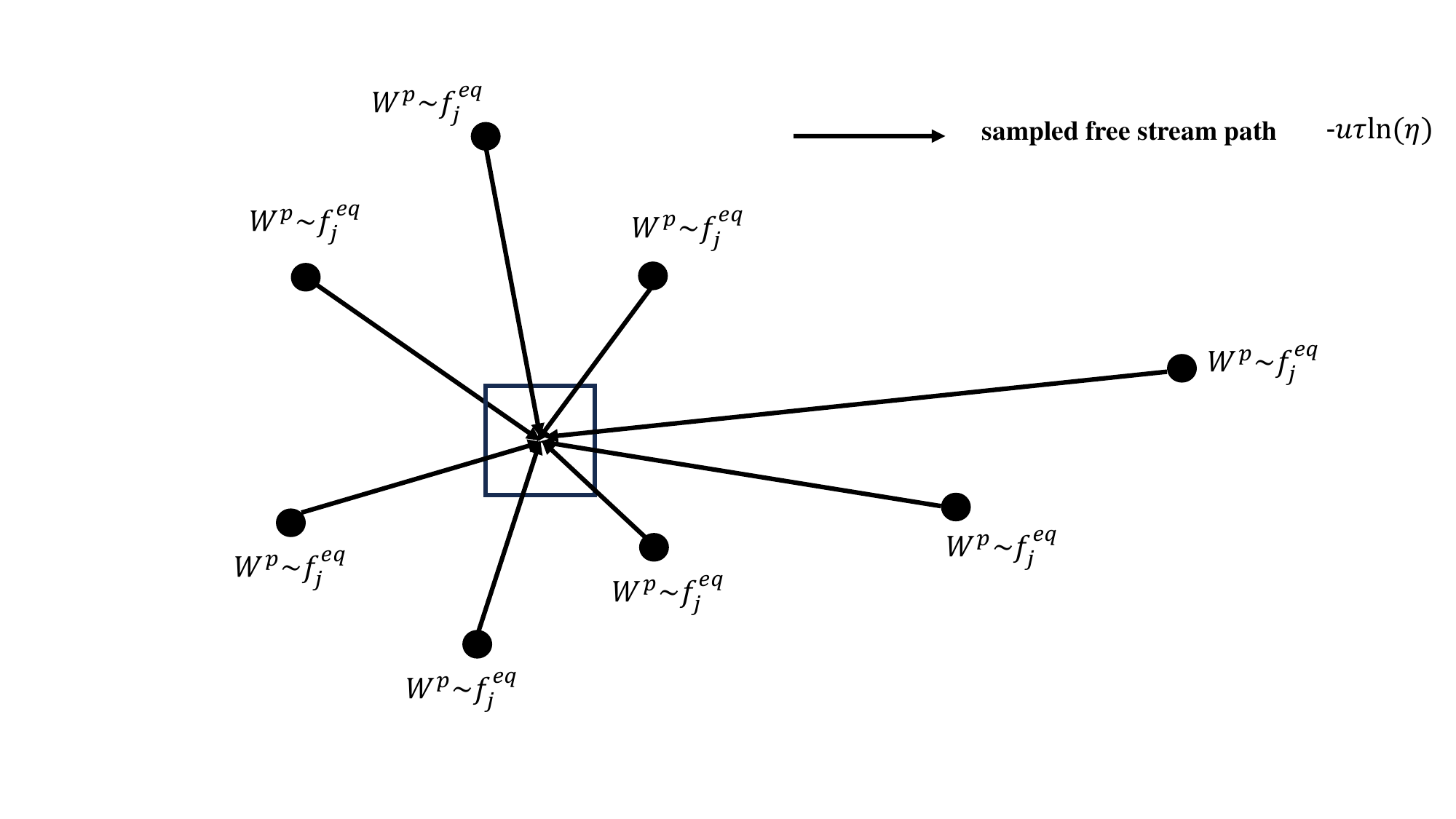}
	\caption{\label{constant-tau-integral-particle}
		Illustrating the integral solution of the particle perspective. }
\end{figure}

Therefore, for a constant value of $\tau$, the steps for one iteration of the steady-state algorithm are:

(1) for each particle in the cell, resampling its distribution, which satisfies the local equilibrium state.

(2) sampling all the particles' free streaming path and moving them to the target cell according to this length.

(3) Update the cell average macro variables by counting the masses, momentum, and energy of the particles in the cell.

\subsection{General cases}
This section will discuss the adjustments and generalizations of the corresponding algorithm when $\tau$ is variable.
To sample the free path length of a particle when $\tau$ is a variable. Here one may employ either analytical or numerical integration techniques to determine the probability density function (PDF), which in turn allows the derivation of the particle's free transport length.
However, the method above can cause issues of spatio-temporal inconsistency. Although the steady-state BGK equation does not incorporate an explicit time dimension, the parameter $\tau$ nevertheless embodies a specific time scale.

This section will use a uniform thermal conduction problem to illustrate the origin and cause of this issue.
The computational domain, initial conditions, and boundary conditions are shown in Fig.~\ref {spatial-time-consistancy-setup}.
\begin{figure}[htp]	\label{spatial-time-consistancy-setup}
	\centering	
    \includegraphics[height=0.45\textwidth]{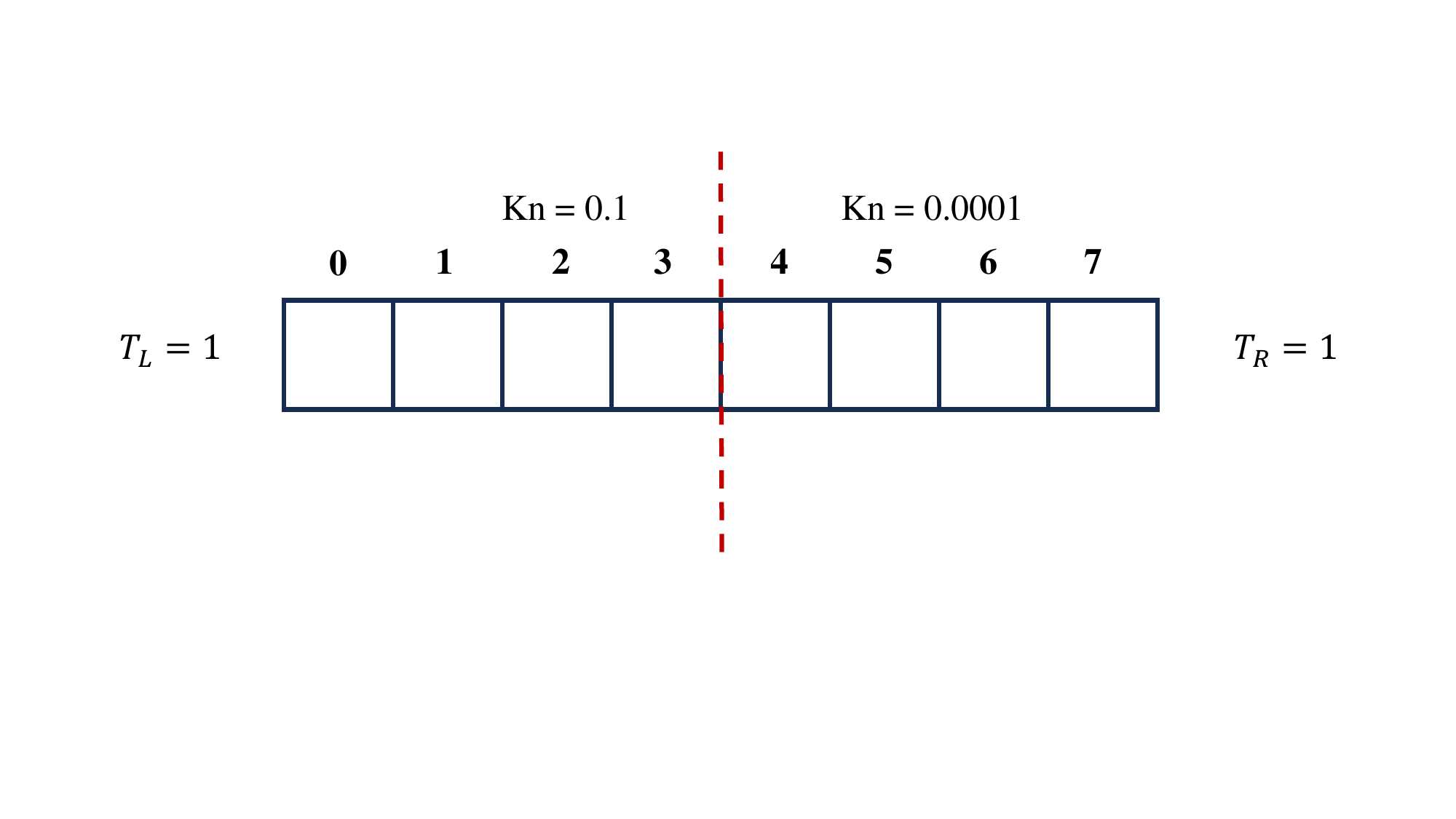}
	\caption{\label{spatial-time-consistancy-setup}
		Computational domain and boundary conditions for the uniform heat conduction case. }
\end{figure}
The grid size is 0.1, and the entire temperature field is initialized to 1, with both the left and right boundary temperatures set to 1. At the middle interface, the Knudsen number (Kn) is 0.1 on the left side and 0.0001 on the right side. With this grid scale and distribution of Kn, particles in the left region move approximately one grid unit per time step, while particles in the right region hardly move. Under these initial and boundary conditions, the entire temperature field should remain at 1.
After sampling N particles in each grid cell, each particle in the entire domain weights (1/N), and the temperature throughout the domain remains at 1, as shown in Fig.~\ref {spatial-time-consistancy-initial}.
\begin{figure}[htp]	\label{spatial-time-consistancy-initial}
	\centering	
    \includegraphics[height=0.45\textwidth]{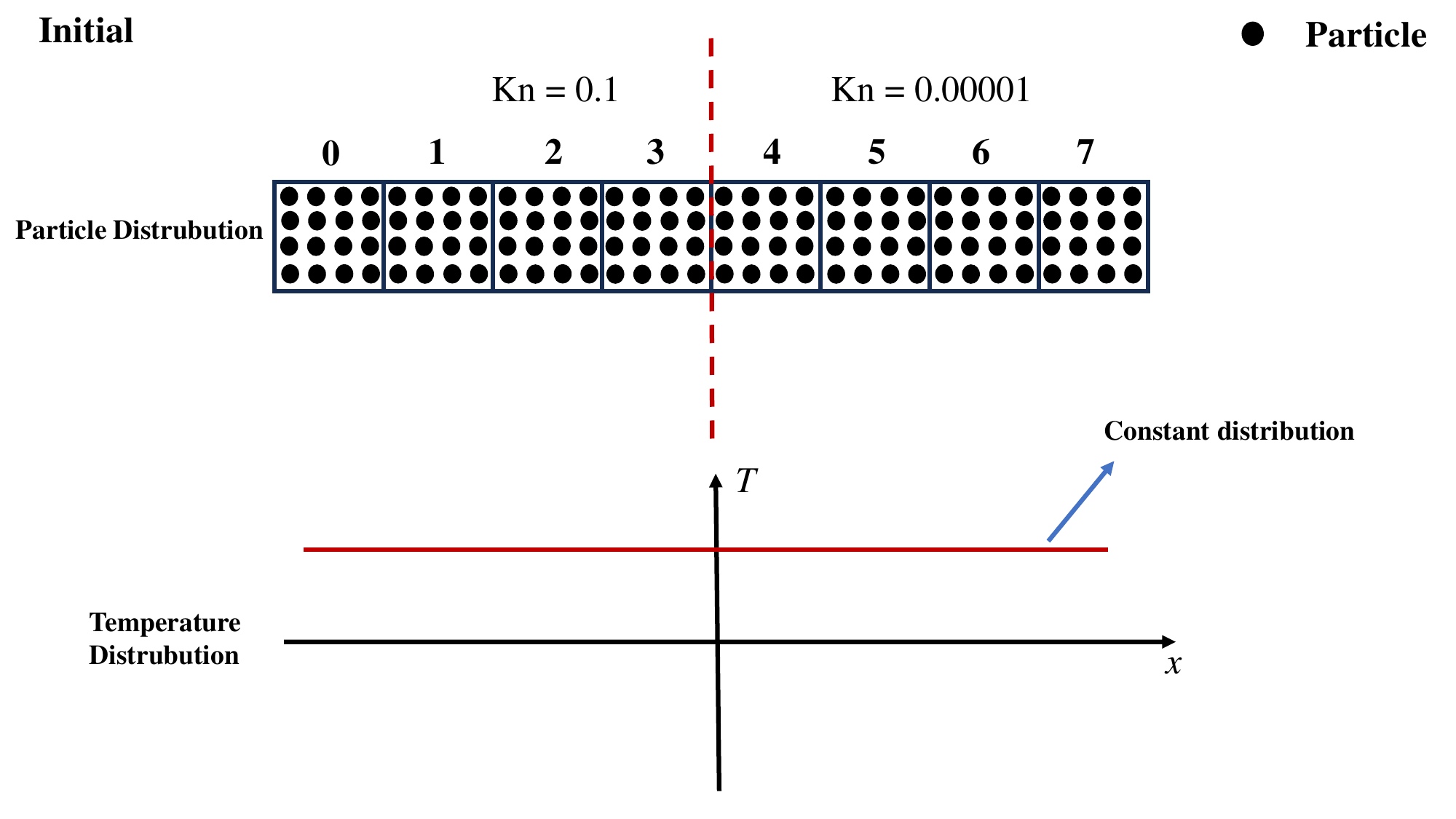}
	\caption{\label{spatial-time-consistancy-initial}
		Initial particle distribution for the uniform heat conduction case. }
\end{figure}
Next, let's focus on the particle distribution in cell three and cell four after one particle movement step. For grid 3, the free path length of the particles is equal to one grid length, and it exchanges an equal number of particles with grid 2; however, about half of the particles move toward grid 4. In contrast, for grid 4, the free path length of the particles is almost zero, and it exchanges an equal number of particles with grid 5, yet virtually no particles move toward grid 3. Therefore, after one iteration, the temperature in grid three will decrease while the temperature in grid four will correspondingly increase.
Fig~.\ref {spatial-time-consistancy-onestep} helps illustrate this transport property.
\begin{figure}[htp]	\label{spatial-time-consistancy-onestep}
	\centering	
    \includegraphics[height=0.45\textwidth]{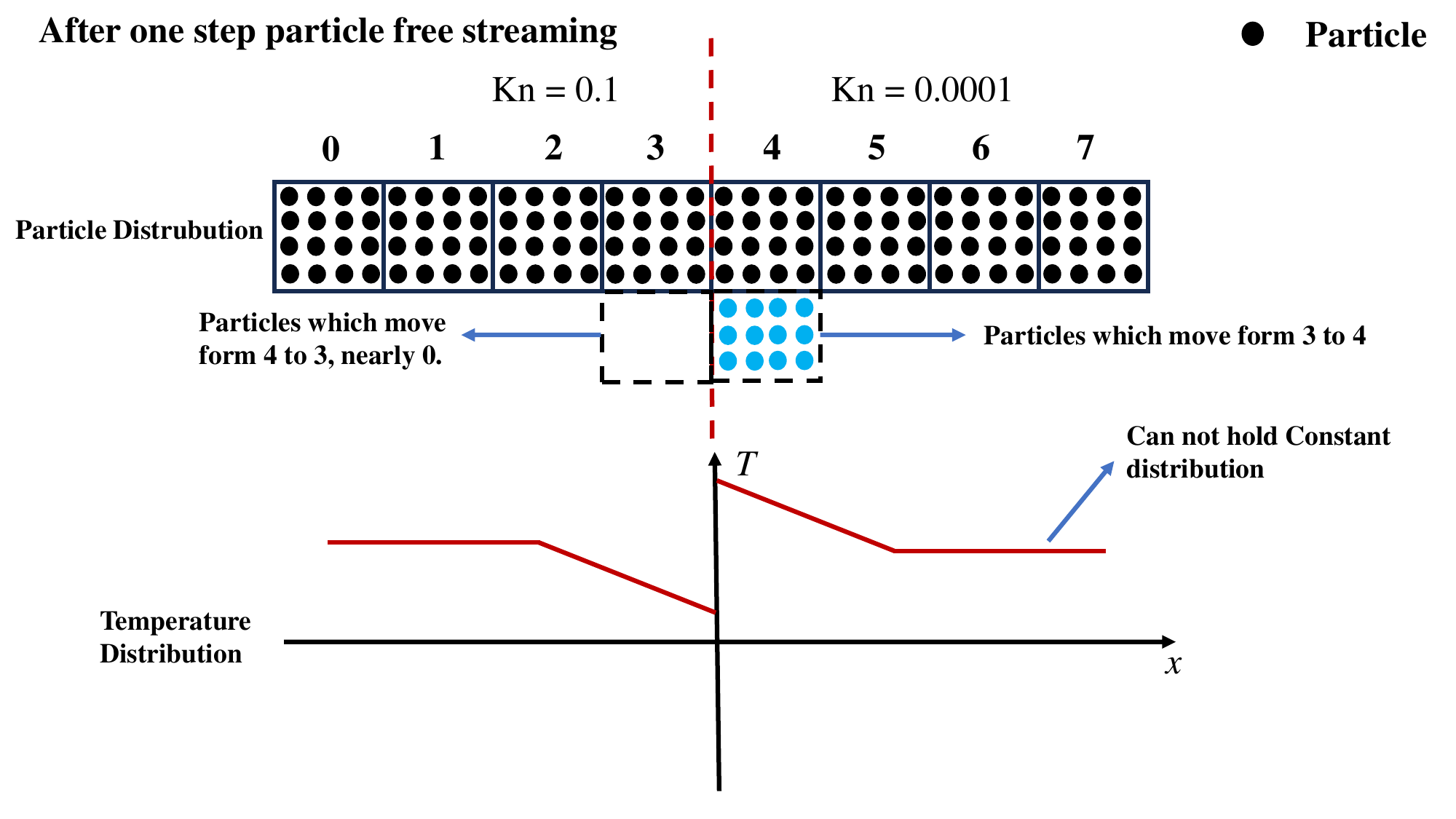}
	\caption{\label{spatial-time-consistancy-onestep}
		Initial particle distribution for the uniform heat conduction case. }
\end{figure}
Moreover, this transport behavior is determined by $\tau$, which depends solely on the spatial coordinate distribution, so this uneven transport does not capture the correct solution in the subsequent iterations.

The fundamental reason for this phenomenon is that the transport times on the left and right sides are not consistent. Particles on the left transport on a timescale corresponding to Kn 0.1, while particles on the right transport on a timescale corresponding to Kn 0.0001, and this discrepancy persists indefinitely as shown in Fig.~\ref{spatial-time-consistancy-transport-property}.
\begin{figure}[htp]	\label{spatial-time-consistancy-transport-property}
	\centering	
    \includegraphics[height=0.45\textwidth]{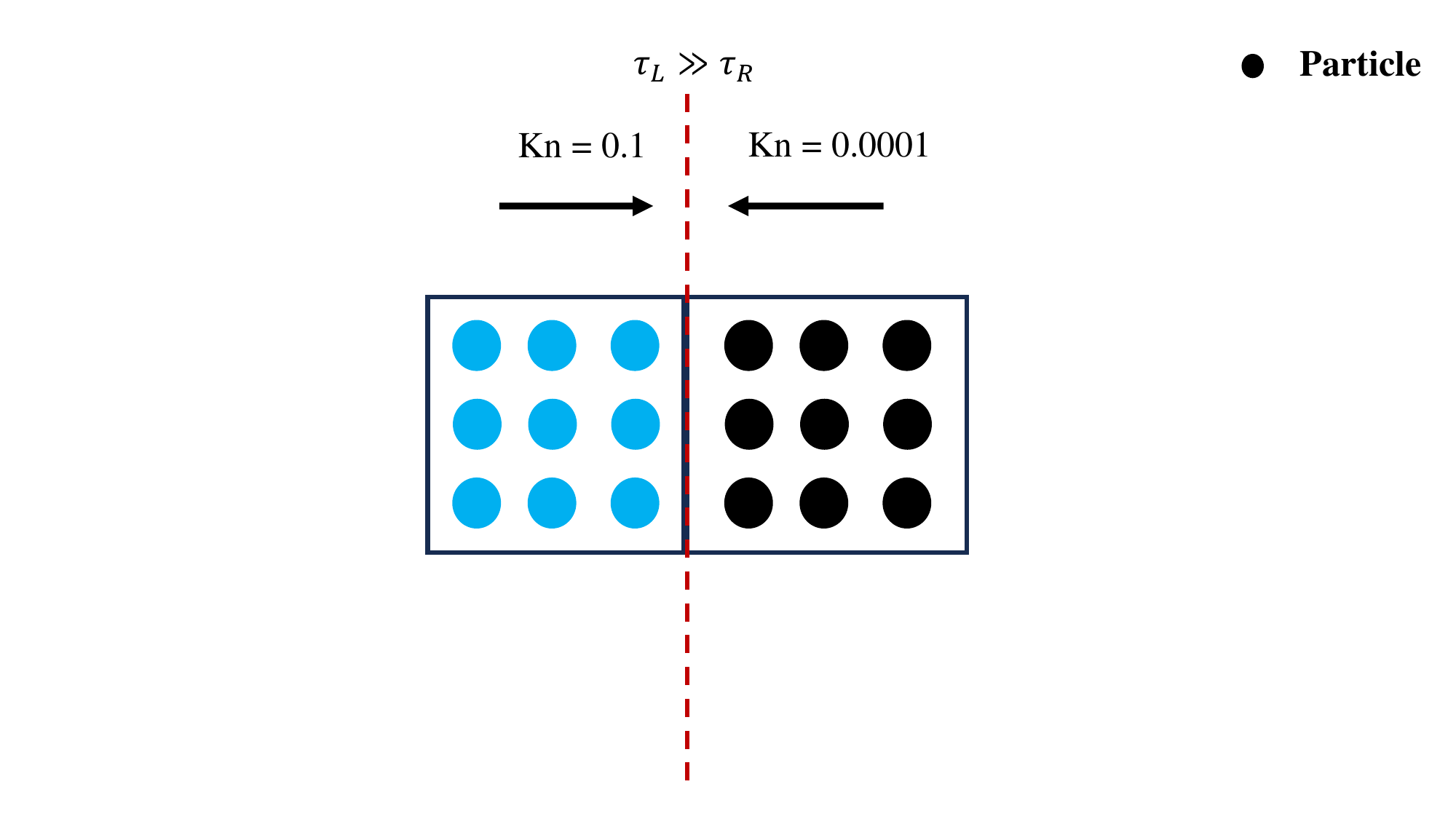}
	\caption{\label{spatial-time-consistancy-transport-property}
		Spatial-time inconsistent transport property. }
\end{figure}

The key to resolve this problem is to define a uniform time scale, so that all particle movements adhere to its constraint.
Moreover, this timescale should resolve the smallest relaxation time in the entire domain, so we select the smallest $\tau$ in the field as the auxiliary timescale denoted as $\tau_{min}$.
After specifying the minimal scale, we can draw inspiration from the null-collision method and the UGKWP method to develop a steady, large-time-step particle method.

For particle sampling, the particle free-path distribution satisfies:
\begin{equation}
-u \tau_{min} \ln (\eta),
\end{equation}
where $u$ is the particle velocity, $\eta$ is a random number in interval $[0,1]$.
Similarly to the UGKWP method, the particle free-transport time $t_f$ is:
\begin{equation}\label{truncated-tf}
    t_f=-\tau_{min}ln\left(\eta\right).
\end{equation}
This differs from the previous method with constant $\tau$.
In the previous method, the true free paths for all particles were sampled, and each particle would undergo a collision after transport, thereby reaching the local equilibrium state. In the current method, the particle free path is first sampled based on $\tau_{min}$, and not all the particles will experience collisions.

To begin with, after free transport, all particles become candidates for particle collisions, and there are two categories of particles: true collisions and null collisions.
This strategy is employed to recover the spatial distribution of $\tau\left(x\right)$.
Here we provide the conditions for null collisions, that is, the acceptance-rejection method. Based on $\tau_{min}$, the probability for the particle's free transport path is:
\begin{equation}
P_c(s)=\frac{1}{u\tau_{min}} e^{\left(-\frac{1}{u\tau_{min}} s\right)}.
\end{equation}
When a candidate collision particle reaches the terminal position x, the ratio between the local true collision frequency $\sigma(x)=\frac{1}{u\tau(x)}$ and the maximum collision frequency $\sigma_{max}=\frac{1}{u\tau_{min}}$ is used to determine whether the candidate collision is a “real collision.” Specifically, the acceptance probability is given by:
\begin{equation}
P_{real}\left(x\right)=\sigma(x) / \sigma_{max}.
\end{equation}
If the randomly sampled value falls within the interval $[0, \sigma(x)/\sigma_{max}]$, then it is considered that a true collision occurs at that position; otherwise, the candidate collision is regarded as a null collision.
In this approach, candidate collision events are generated at a uniform rate of $\sigma_{max}$, and the local true collision probability is recovered through the acceptance-rejection mechanism.
Next, we will introduce how this method reconstructs the probability density of the free path length for true collisions.

Let's denote the free-transport distance from the location of the last collision to the position where the first candidate collision is accepted as a true collision after multiple candidate collisions as $s$.
Let $f(s)$ represent the probability density that the particle experiences a true collision in the interval from $x$ to $x + s$.
One can think of building $f(s)$ by summing over all possible scenarios where the particle undergoes $n$ candidate collisions that are all rejected and then, at distance $s$, a candidate collision is accepted as a true collision. Here $n$ can be $0, 1, 2, 3,...$.

When $n=0$, the first candidate collision is accepted; the candidate collision event at $s$ is a true collision. The candidate collision-free-path length $s$ follows an exponential distribution given by $P_c(s) = \sigma_{max} e^{(–\sigma_{max} s)}$, and the probability that this candidate collision at position $x+s$ is accepted as a true collision is $\sigma(x+s)/\sigma_{max}$. Thus, the corresponding probability of $n=0$ is:
\begin{equation}
f_0(s)=  P_c(s) \left[\frac{\sigma(x+s)}{\sigma_{max}}\right]
 = \sigma_{max} e^{\left(-\sigma_{max} s\right)}\left[\frac{\sigma(x+s)}{\sigma_{max}}\right]
 = \sigma(x+s) e^{ \left(-\sigma_{max} s\right)}.
\end{equation}

When $n=1$, for the first candidate collision occurring at $s_1$, where $s_1 \in[0, s]$, the probability $P_{nc}(s_1)$ that it is a null collision is:
\begin{equation}
P_{nc}(s_1)=P_c(s_1)\left[1-\sigma\left(x+s_1\right) / \sigma_{m a x}\right].
\end{equation}
The second candidate collision, occurring a distance $s_2 = s–s_1$ after the first candidate collision, has a probability given by:
\begin{equation}
P_c(s_2)=\sigma_{max} e^{ \left[-\sigma_{max} \left(s-s_1\right)\right]}.
\end{equation}
And at position $x+s$ it is accepted as a true collision with probability $\sigma(x+s)/\sigma_{max}$. After integrating over $s_1$, we obtain:
\begin{equation}
    \begin{aligned}
        f_1(s)=&\int_0^sP_c(s_1)\left[1-\sigma\left(x+s_1\right) / \sigma_{m a x}\right] \cdot P_c(s_2)\left[\sigma(x+s)/\sigma_{max}\right]ds_1\\
        =& \sigma(x+s)e^{-\sigma_{max}s}\int_0^s\sigma_{max}-\sigma(x+s_1)ds_1.
    \end{aligned}
\end{equation}

Similarly $f_n(s)$ can be obtained:
\begin{equation}
    f_n(s)=\sigma(x+s)e^{-\sigma_{max}s}\phi_n(s),
\end{equation}
where $\phi_n(s)$ is the n-fold convolution integral contributed by n times null collisions:
\begin{equation}
\phi_n(s)=(1 / \mathrm{n}!)\left[\int_0^{\mathrm{s}}\left[\sigma_{max} -\sigma(x+u)\right] \mathrm{du}\right]^{\mathrm{n}}.
\end{equation}

Finally, summing over all n, we obtain the overall probability density for a true collision:
\begin{equation}
\begin{aligned}
        f(s)=& \sum_{n=0}^{\infty} f_n(s) \\
        =& \sigma(s)e^{-\sigma_{max}s}
        \sum_{n=0}^{\infty}(1 / \mathrm{n}!)\left[\int_0^{\mathrm{s}}\left[\sigma_{max} -\sigma(u)\right] \mathrm{du}\right]^{\mathrm{n}} \\
        =& \sigma(s)e^{-\sigma_{max}s}
        e^{ \left[\int_0^s\left(\sigma_{max} -\sigma(u)\right) d u\right]} \\
        =& \sigma(s)e^{-\int_0^s\sigma(u)du}.
\end{aligned}
\end{equation}
The "accept-reject" method has now recovered the PDF of the particle free stream path.

Based on the preceding discussion, we now distinguish two classes of particles: null-collision particles and real collision particles. Both null-collision and real collision particles are considered generalized collision particles; however, the distinction lies in their roles. Null-collision particles contribute solely to the statistical evaluation of macroscopic quantities without adhering to the local equilibrium state. In contrast, real collision particles participate in the macroscopic statistical measures and conform to the local equilibrium distribution.

It is important to emphasize that, unlike the previous method in which $\tau$ was treated as a constant where the equilibrium state of the particles was determined by the total macroscopic quantities of the cell because every particle contributing to the cell’s macroscopic quantities underwent a true collision. The current method, similar to UGKP, determines the particle equilibrium state by $W^h$, which is the macroscopic quantity arising from the summation of all real collision particles.
Then, the cell total macroscopic variable $W$ and wave macroscopic variable $W^h$ is:
\begin{equation}
    W=\sum w^{nc}+\sum w^{rc},W^h=\sum w^{rc},
\end{equation}
where $w^{nc}$ is the null collision particle's quantity and $w^{rc}$ is the real collision particle's quantity.

The reason for re-sampling the equilibrium state of real collision particles from $W^h$ instead of $W$ is that the macroscopic quantity contributed by real collision particles is $W^h$, not $W$. Furthermore, sampling from $W$ would lead to a violation of the conservation of total weight.
Fig ~.\ref {spatial-time-consistancy-particle-classify} helps illustrate the two types of particles.
\begin{figure}[htp]	\label{spatial-time-consistancy-particle-classify}
	\centering	
    \includegraphics[height=0.50\textwidth]{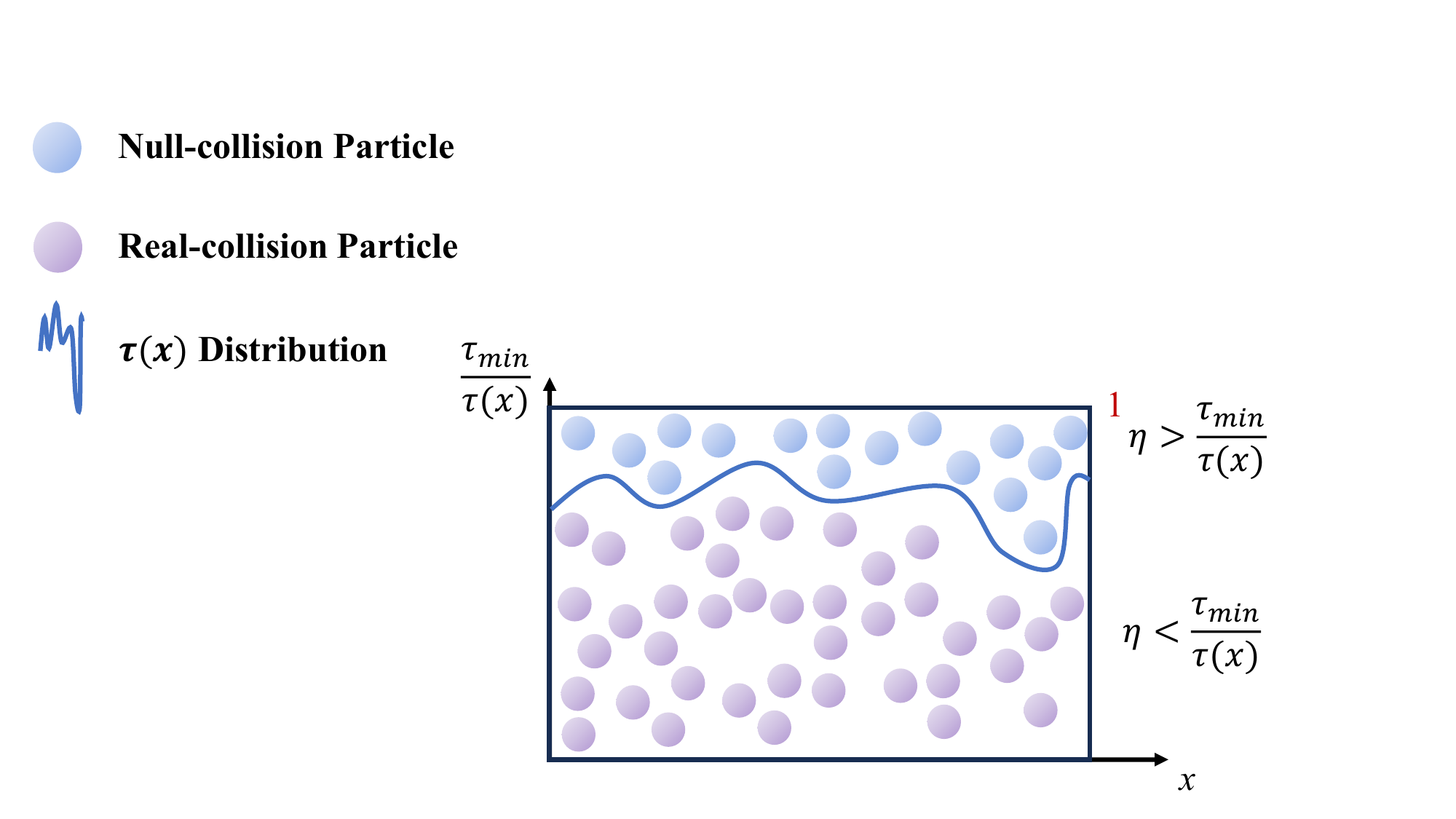}
	\caption{\label{spatial-time-consistancy-particle-classify}
		Two types of particles in space. }
\end{figure}
Thus far, this paper summarizes the steps of the steady-state acceleration method for variable $\tau$ as follows:

(1) If it is the first calculation step, sample N particles in each cell, with each particle following the local equilibrium distribution ($w^{ini}=\frac{W}{N}$, $w^{ini}$ is the initial energy for each particle) and its free path being sampled; if it is not the first step, skip this process.

(2) If it is not the first calculation step, then in each cell, for real collision particles, resample their equilibrium distribution, which is determined by the current cell value $W^h$ ($w^{rc}=\frac{W^h}{N_{rc}}$, $N_{rc}$ is the real collision particle number in this cell), and also resample their free path.

(3) Based on the particle's free path, transport the particle to its destination.

(4) Upon reaching its destination, the previously described acceptance–rejection method is used to determine whether it corresponds to a real collision or a null collision.

(5) Accumulate the total energy $W$, contributed by the null collision and real collision particles to each cell.
Accumulate the energy of the wave component $W^h$ contributed by the real collision particles to each cell. And count the number of real collision particles within each cell, denoted as $N_{rc}$.

(6) For null collision particles, resample the free path for the next step.

\subsection{Macro prediction acceleration for
steady state kinetic particle method}
At small Knudsen numbers, for instance, Kn = 0.001, the current method results in an extremely small particle mean free path, which necessitates a substantial number of iterations to reach the final steady-state solution.
Meanwhile, in the small Knudsen number limit, the macroscopic equations corresponding to the BGK model are in excellent agreement with the exact solution.
Inspired by this insight, we can leverage the macroscopic equations to rapidly evolve the flow or temperature fields, thereby providing an estimation of the particles' equilibrium state for the next step.
This section will illustrate how to construct the macro prediction equation and introduce the iterative mechanism between the macroscopic and microscopic equations.

Taking the zeroth moment of the steady-state BGK equation yields the energy conservation equation:
\begin{equation} \label{macro-q-eqn}
    \nabla \cdot \boldsymbol{q} =0 ,
\end{equation}
where $\boldsymbol{q}$ is the heat flux.
This equation is universally valid for all Knudsen
numbers at steady state.
Since the relationship between $\boldsymbol{q}$ and $T$ is unknown, Eq ~.\eqref{macro-q-eqn} cannot be solved directly.
Therefore, inexact Newton iteration can be employed.
The residual formulation of Eq~.\eqref{macro-q-eqn} can be written as:
\begin{equation} \label{macro-q-res}
    \nabla \cdot \boldsymbol{q}=Q\left(T\right) = R,
\end{equation}
where $R$ is the residual.
As the residual vanishes, Eq~.\eqref{macro-q-eqn}
is satisfied.
Therefore, we introduce a linear operator $\tilde{Q}$ to find an increment of $T$ that might diminish the residual. The specific expression of $\tilde{Q}$ adopted in this study is given as follows:
\begin{equation}\label{linear-q}
\tilde{Q}(\Delta T)=\nabla \cdot\left(-\kappa \nabla \Delta T\right).
\end{equation}
Then, combining Eq~.\eqref{macro-q-res} and Eq~.\eqref{linear-q} we can construct a delta form for temperature:
\begin{equation}
    \nabla \cdot \boldsymbol{q} = \nabla \cdot \left[\kappa \nabla \left(\Delta T\right)\right].
\end{equation}
The heat flux divergence also represents the net heat inflow into the cell, which equates to the cell’s energy increment. This energy increment can be obtained by subtracting the cell’s total real collision particle energy from the previous and current steps:
\begin{equation}
    \nabla \cdot \boldsymbol{q} = \sum w^{n+1} - \sum w^{n}.
\end{equation}
Then, macro energy can be updated by $E^{n+1}=E^{n}+C_v\Delta T$.
Thus, the new equilibrium state for real collision particles at the next iteration can be sampled from this predicted $E^{n+1}$.

\section{Numerical Test}
This section tests a series of benchmark heat transfer cases to validate the accuracy and efficiency of the proposed multiscale method for solving the phonon BTE.
The results obtained by the present method will be
compared with the data predicted by the DUGKS \cite{zhang2019dugksphonon} and UGKS.
Moreover, $C_v$ and $\boldsymbol{|V_g|}$ is 1 in 1D and 2D cases  if not specified.

For all the cases, the speedup ratio is computed by:
\begin{equation}
    R_s=\frac{N_{wp}}{N_{current}}R_e,
\end{equation}
where $R_s$ is the final speedup ratio, $N_{wp}$ is the number of steps required for the explicit UGKWP to reach a steady state. $N_{current}$ is the number of steps required for the current method to reach a steady state, $R_e$ is the CPU time speedup ratio for each step.
\subsection{One-dimensional heat conduction in a dielectric film}
In this section, the one-dimensional heat conduction in a dielectric film with a thickness of $L = 1$ is simulated, as illustrated in Fig~.\ref{1d-heat}.

\begin{figure}[htp]	\label{1d-heat}
	\centering	
	\includegraphics[height=0.45\textwidth]{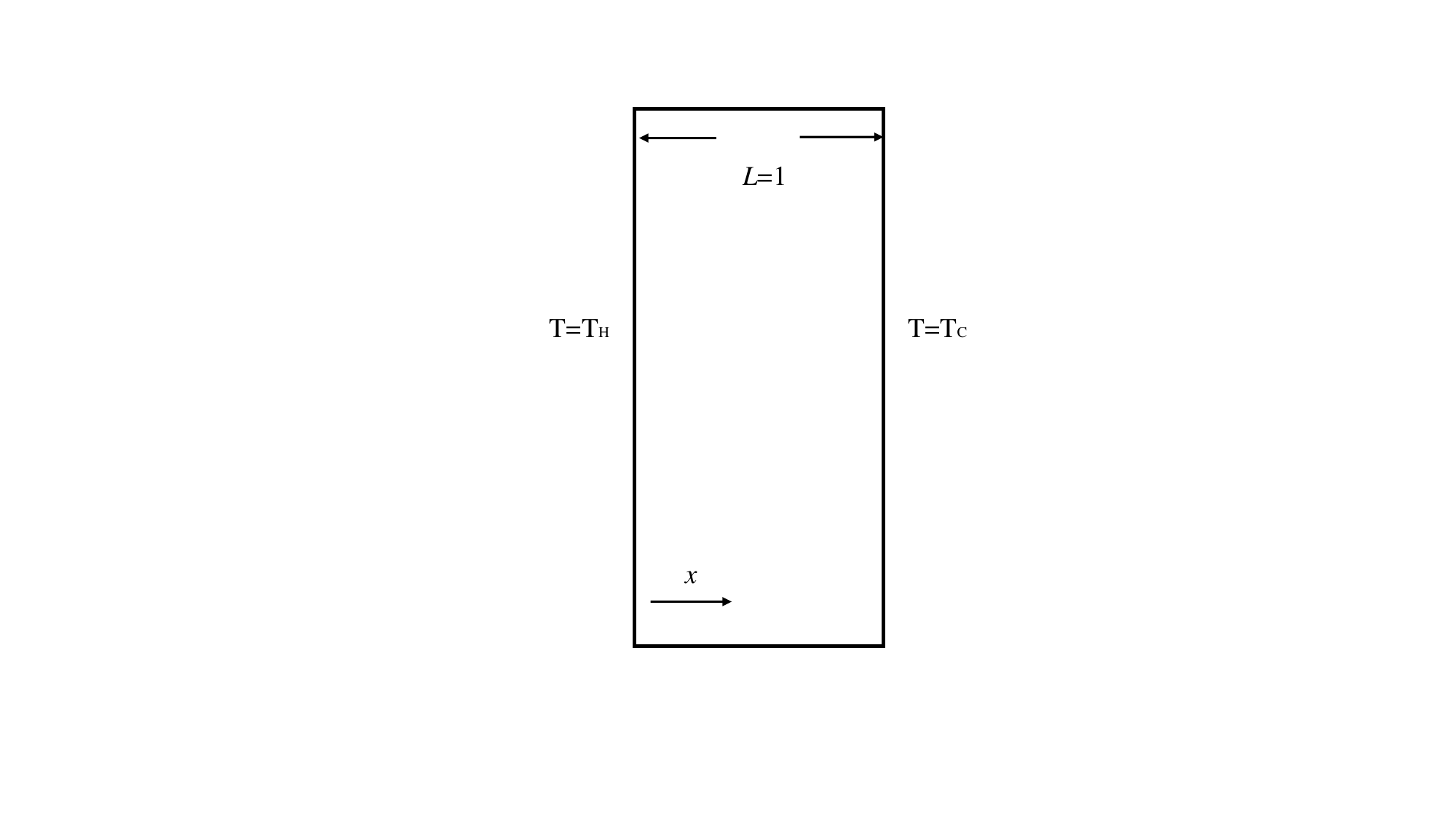}
	\caption{\label{1d-heat}
		A schematic diagram of one-dimensional heat conduction in a dielectric film.}
\end{figure}
At the left boundary $(x = 0)$, an isothermal high-temperature boundary is imposed with a temperature of $T_H$, while at the right boundary $(x = L)$, an isothermal low-temperature boundary is imposed with a temperature of $T_L$.
An analytical solution can be obtained in \cite{heaslet1965radiative-1d-ana-1, majumdar1993microscale-1d-ana-2}.

In this test case, the one-dimensional heat conduction problem is computed for Knudsen numbers of 10.0, 1.0, 0.1, 0.01, and 0.001. The corresponding reference particle numbers are 8000 to balance computational cost and statistical noise.
The one-dimensional computational domain is discretized into 100 uniform cells, and the CFL number is set to 0.5 for the UGKWP method to compare the CPU costs with the current method.
The results indicate that the computational outcomes of the current method are in excellent agreement with the analytical solutions.

Fig~.\ref{1d-heat-result} illustrates a comparison among the computed results of the current method and the analytical solution under different Kn regimes,

\begin{figure}[htp]	\label{1d-heat-result}
	\centering	

    \includegraphics[height=0.40\textwidth]{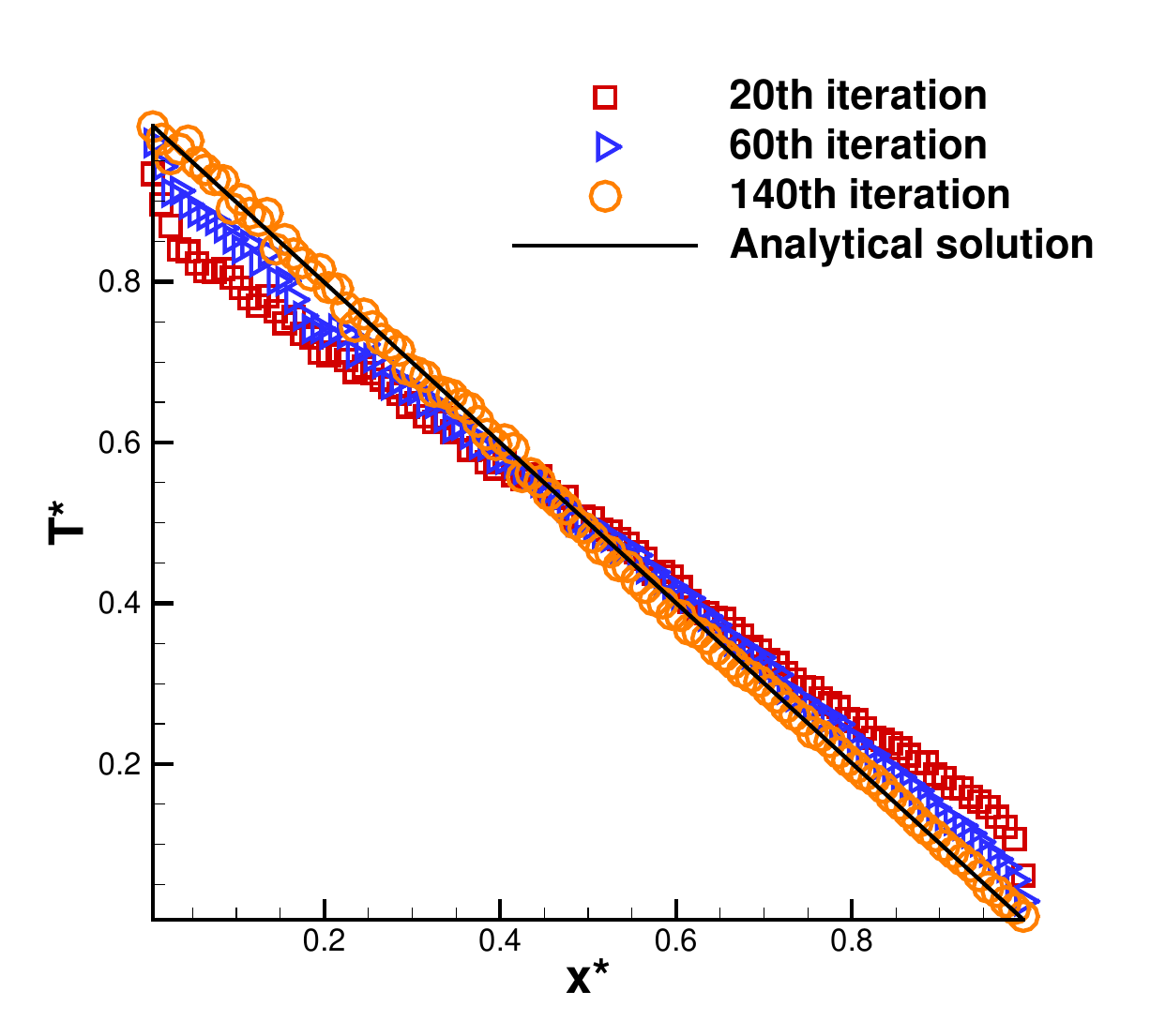}
	\includegraphics[height=0.40\textwidth]{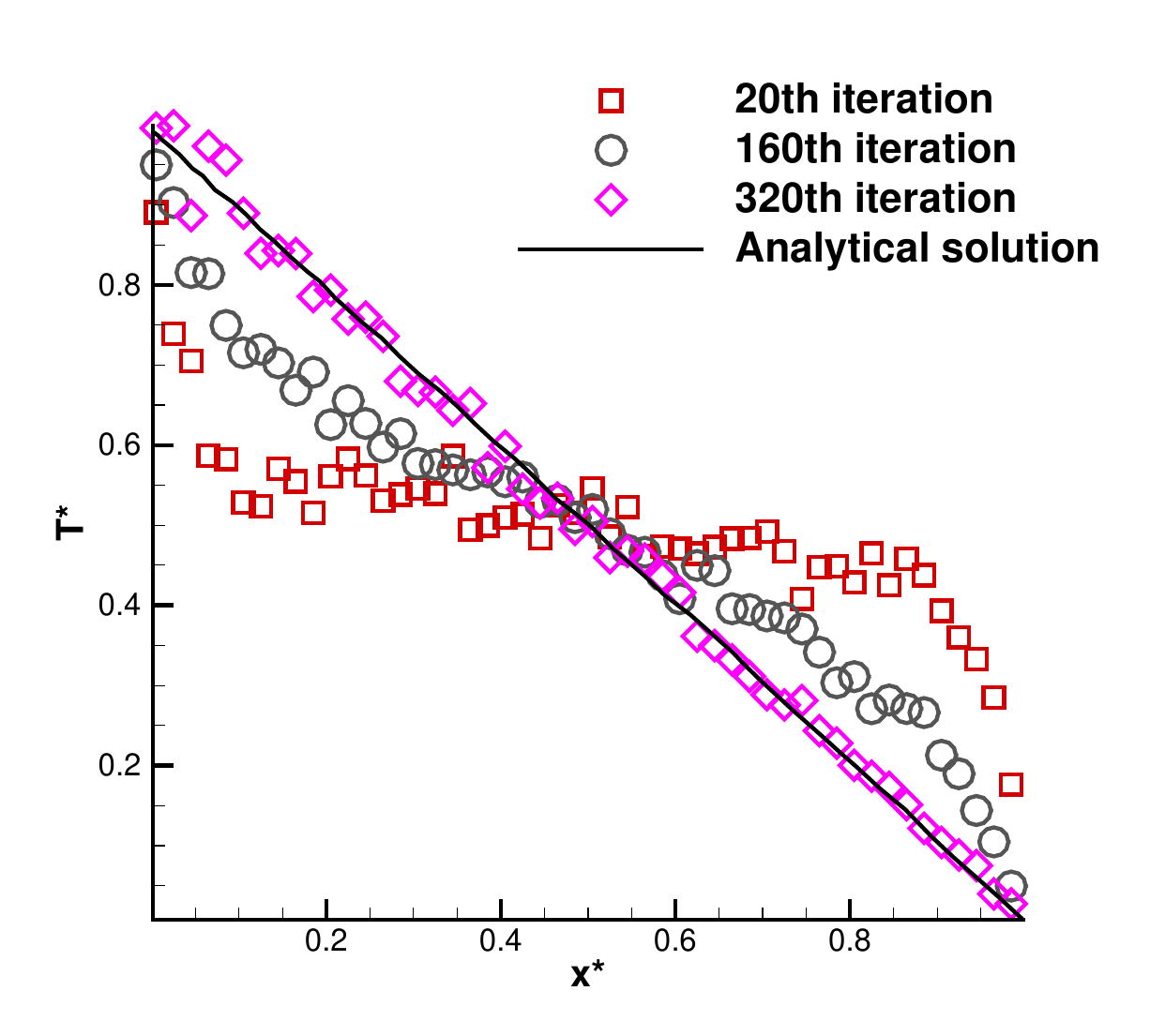}
        \includegraphics[height=0.40\textwidth]{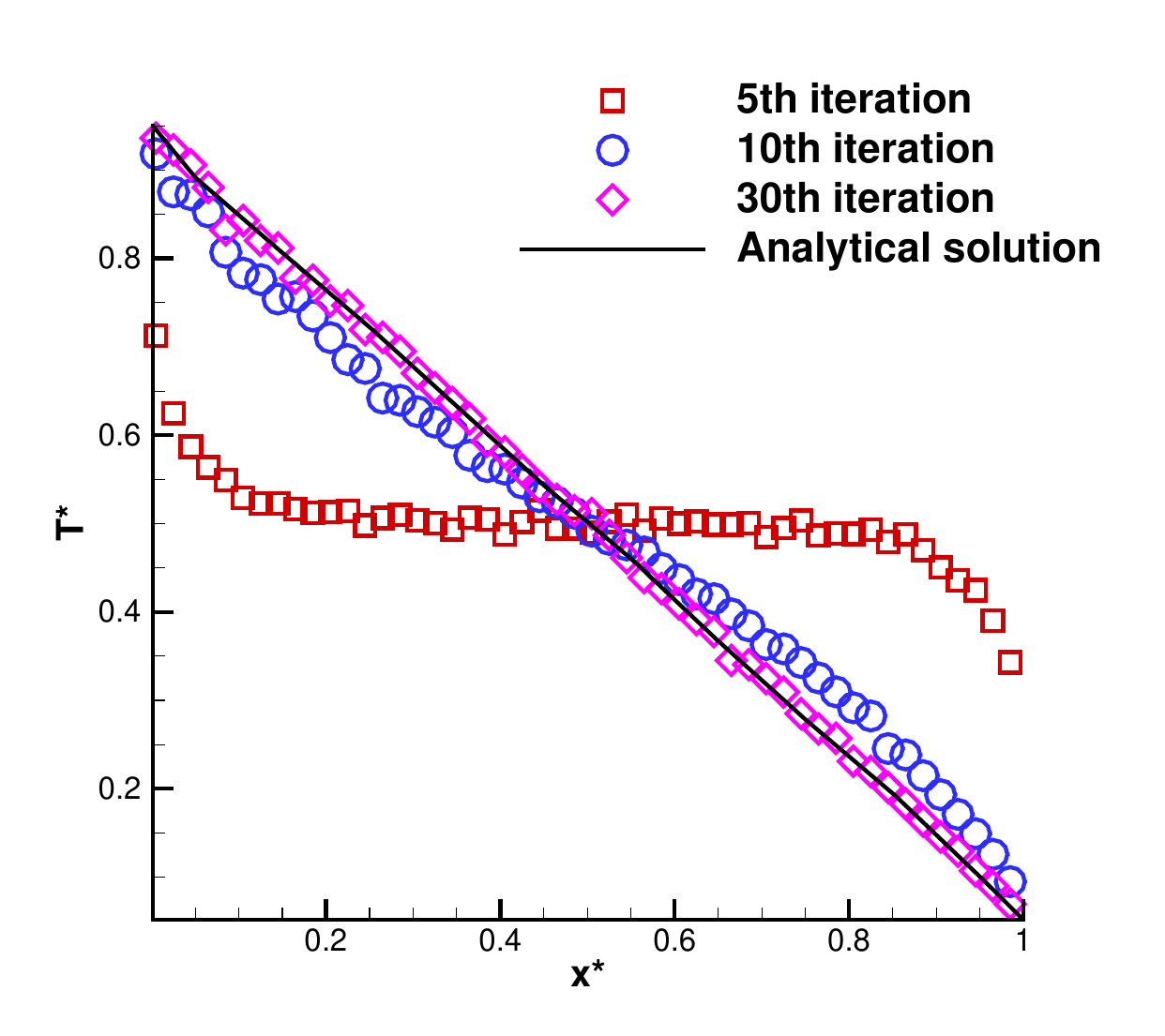}
        \includegraphics[height=0.40\textwidth]{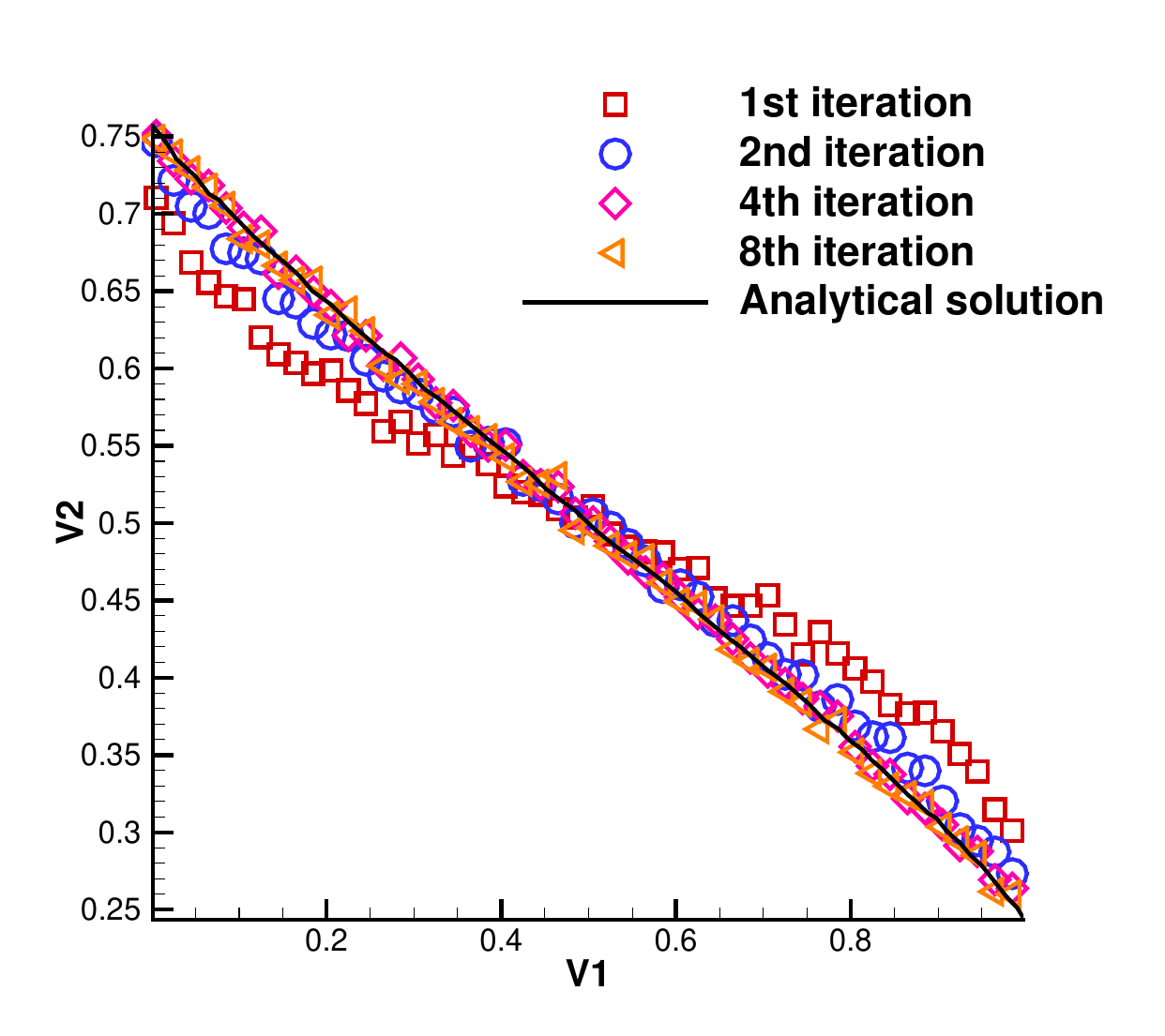}
        \includegraphics[height=0.40\textwidth]{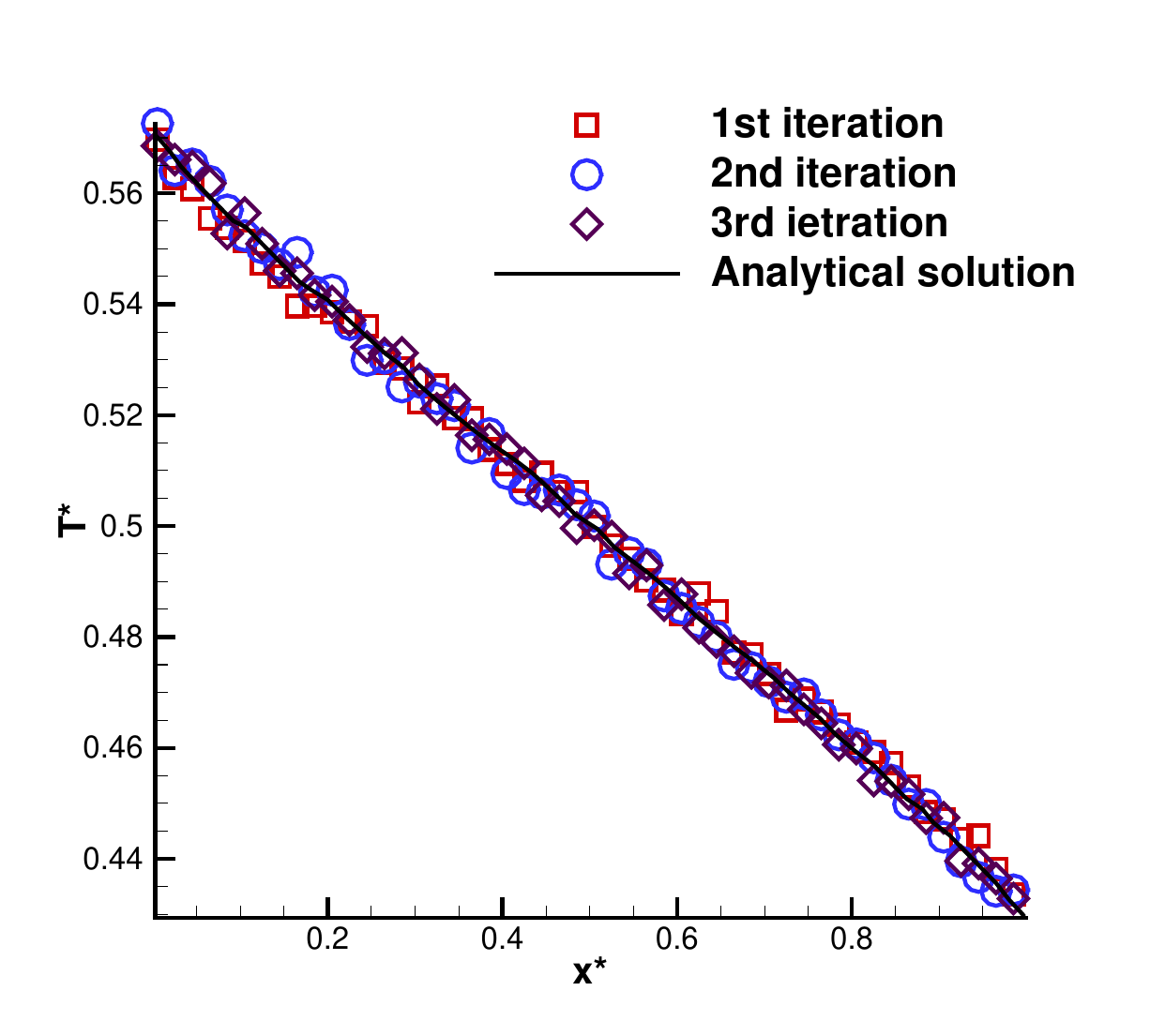}

	\caption{\label{1d-heat-result}
		Comparison of 1D heat conduction results across a film, from Kn = 0.001 to Kn = 10.0. }
\end{figure}

Table~.\ref{1st-table} compares the iteration number and computational time between the UGKWP method and the current method when approaching to the steady state.

\begin{table}[htp]
	\small
	\begin{center}
		\def\temptablewidth{1.0\textwidth}
		{\rule{\temptablewidth}{1pt}}
		\begin{tabular*}{\temptablewidth}{@{\extracolsep{\fill}}c|c|c|c|c|c}
			Method & Kn = 10.0 &  Kn = 1.0 & Kn = 0.1 &  Kn = 0.01 & Kn = 0.001\\
			\hline
			UGKWP & 1200 & 800 & 800 & 3200 & 24000 \\ 	
			Current & 3 & 8 & 40 & 320 & 140 \\ 	
                Speedup & 1600 & 400 & 80 & 40 & 680 \\ 	
		\end{tabular*}
		{\rule{\temptablewidth}{0.1pt}}
	\end{center}
	\vspace{-4mm} \caption{\label{1st-table} Comparison of the computational costs of one-dimensional heat conduction in a dielectric film between the UGKWP method and the current method. (Each step in the current method is about 4 times faster than the original explicit UGKWP method, which indicates that $R_e = 4$.)}
\end{table}

Based on computational costs, the current method significantly reduces the number of iterations required to reach the steady state compared to the explicit UGKWP method.
Moreover, the current approach does not physically add or remove particles from the grid. Instead, the collision process is implemented in the subsequent iteration step by enforcing that the particles conform to the local equilibrium distribution. So, the particle count remains fixed.
These two aspects enable using a simple contiguous array for particle storage, eliminating the need for a complex doubly-linked list structure, which is used in the UGKWP method in this paper.
Consequently, the CPU time per iteration in the current method is greatly reduced compared to that of the UGKWP method.

As previously mentioned, the current method demonstrates the highest efficiency in large Knudsen number regimes. In small Knudsen number cases, macroscopic equations can be introduced to estimate the equilibrium state, thereby speeding up the acceleration. The results indicate that this coupling strategy achieves a favorable acceleration ratio.

\subsection{One-dimensional heat conduction with discontinuous $\tau$ }

In the previous section, we demonstrated the current method's computational accuracy and accelerated convergence under constant $\tau$.
This section will evaluate the current method's computational accuracy and accelerated convergence when $\tau$ varies with the spatial coordinate to validate the effectiveness of the proposed space–time consistent treatment.
This section will examine a heat conduction problem with a piecewise discontinuous $\tau$. For $x < 0.5L$, $\tau$ is set to 10.0; for $x > 0.5L$, $\tau$ is set to 0.1. In other words, at the interface in the middle, the values of $\tau$ differ by 100.

At the left boundary $(x = 0)$, an isothermal high-temperature boundary is imposed with a temperature of $T_H$, while at the right boundary $(x = L)$, an isothermal low-temperature boundary is imposed with a temperature of $T_L$.
The reference particle number is 8000 to balance computational efficiency with statistical noise.
The one-dimensional computational domain is discretized into 100 uniform cells, and the CFL number is set to 0.5 for the UGKWP method to compare the CPU costs with the current method.
In this section, we also computed the case where $\tau$ on the left and right sides differ by a factor of 10. Specifically, $\tau$ is ten on the left side and one on the right side.
For convenience, we refer to the configuration with a 100-fold difference in Knudsen numbers as Case A, and the one with a 10-fold difference as Case B.

The computational results, as shown in Fig~.\ref{1d-heat-result-dis}, indicate that when using the UGKS method as the reference solution, the current method is in very good agreement with UGKS.

\begin{figure}[htp]	\label{1d-heat-result-dis}
	\centering	
	\includegraphics[height=0.40\textwidth]{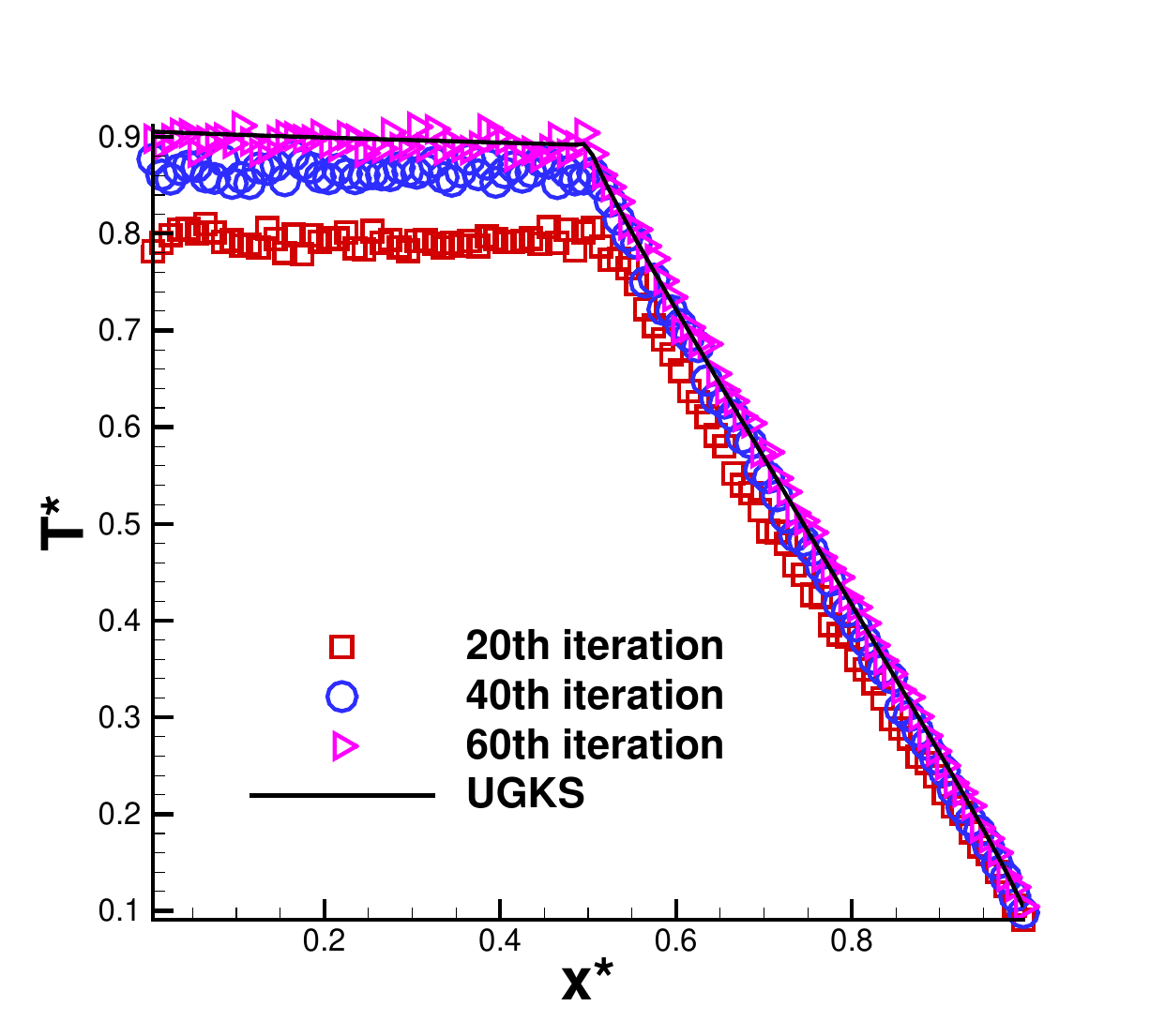}
        \includegraphics[height=0.40\textwidth]{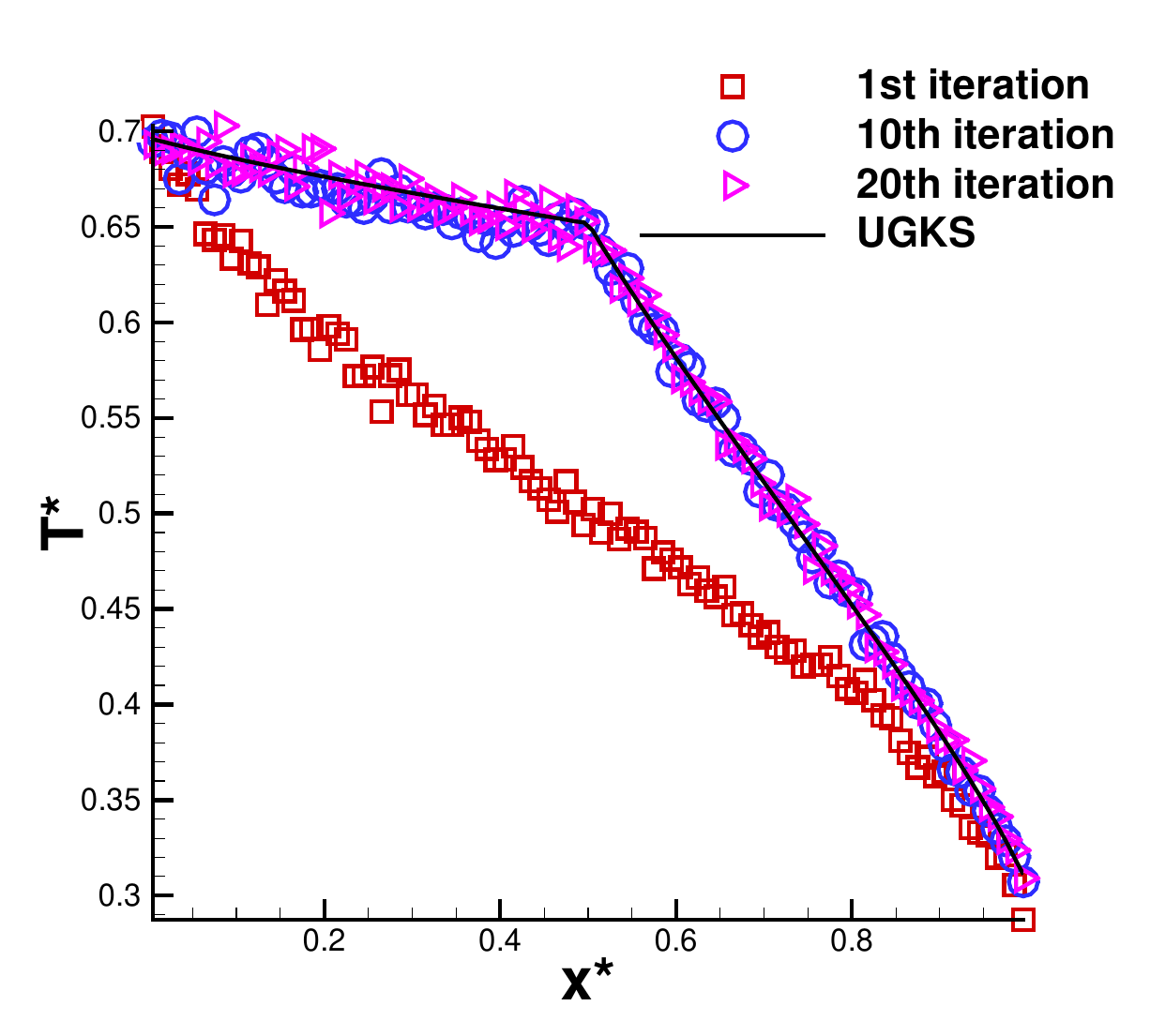}

	\caption{\label{1d-heat-result-dis}
		Comparison of 1D heat conduction results across a film, discontinuous $\tau$ case. Left: $\text{Kn}_{L}=10.0$, $\text{Kn}_{R}=0.1$. Right: $\text{Kn}_{L}=10.0$, $\text{Kn}_{R}=1.0$ }
\end{figure}

Table~.\ref{2nd-table} compares the iteration count and computational time between the UGKWP method and the current method when approaching to the steady state solution.

\begin{table}[htp]
	\small
	\begin{center}
		\def\temptablewidth{1.0\textwidth}
		{\rule{\temptablewidth}{1pt}}
		\begin{tabular*}{\temptablewidth}{@{\extracolsep{\fill}}c|c|c}
			Method & Case A &  Case B \\
			\hline
			UGKWP & 1500 & 800  \\ 	
			Current & 60 & 20  \\ 	
                Speedup & 100 & 160  \\ 	
		\end{tabular*}
		{\rule{\temptablewidth}{0.1pt}}
	\end{center}
	\vspace{-4mm} \caption{\label{2nd-table} Comparison of the computational costs of one-dimensional heat conduction in a dielectric film between the UGKWP method and the current method, discontinuous $\tau$ case. (Each step in the current method is about 4 times faster than the original explicit UGKWP method, which indicates that $R_e = 4$.)}
\end{table}

The efficiency comparison shows that when $\tau$ is discontinuously distributed in space, the current method still achieves a two-order-of-magnitude improvement in convergence efficiency compared to the UGKWP method. In this case, the convergence efficiency is not as high as that with constant $\tau$; the reason is that we establish a globally uniform time scale based on the smallest $\tau$ in the domain.
Therefore, the free path of a particle in each iteration step is shorter than the free path sampled with a constant $\tau$, which consequently necessitates more iterations to reach the steady state.

\subsection{One-dimensional multi-scale cases}
This section designs a one-dimensional multiscale heat conduction case to further demonstrate the proposed method's capability for capturing multiscale non-equilibrium phenomena in phonon transport.
The computational setup and boundary conditions are identical to those in the first test case, except that the relaxation time is now a spatially varying function, as detailed below.
\begin{equation}
\tau(x)=10^{|A_1 \sin (2 \pi x / L)|-A_2}.
\end{equation}
Here, $A_1$ and $A_2$ are adjustable constants. When $A_1$ is 2.0 and $A_2$ is -1.0, the value of $\tau$ spans the range $[10^{-1}, 10^{1}]$. This indicates that the Knudsen number varies by two orders of magnitude over the entire spatial domain, posing a significant challenge for multiscale methods. Consequently, this example is an excellent test case for evaluating the performance of the present method.
Initially, the temperature throughout the entire computational domain is uniformly set to $0.5 (T_H + T_c)$, representing the arithmetic mean of the high and low temperature values. The particle reference sampling number is 8000, and the grid number is 100.
The result is shown in Fig~.\ref{multiscale-1d-heat-result}, illustrating that the UGKWP method can automatically recover the heat transfer physics in different scales.

\begin{figure}[htp]	\label{multiscale-1d-heat-result}
	\centering	
	\includegraphics[height=0.40\textwidth]{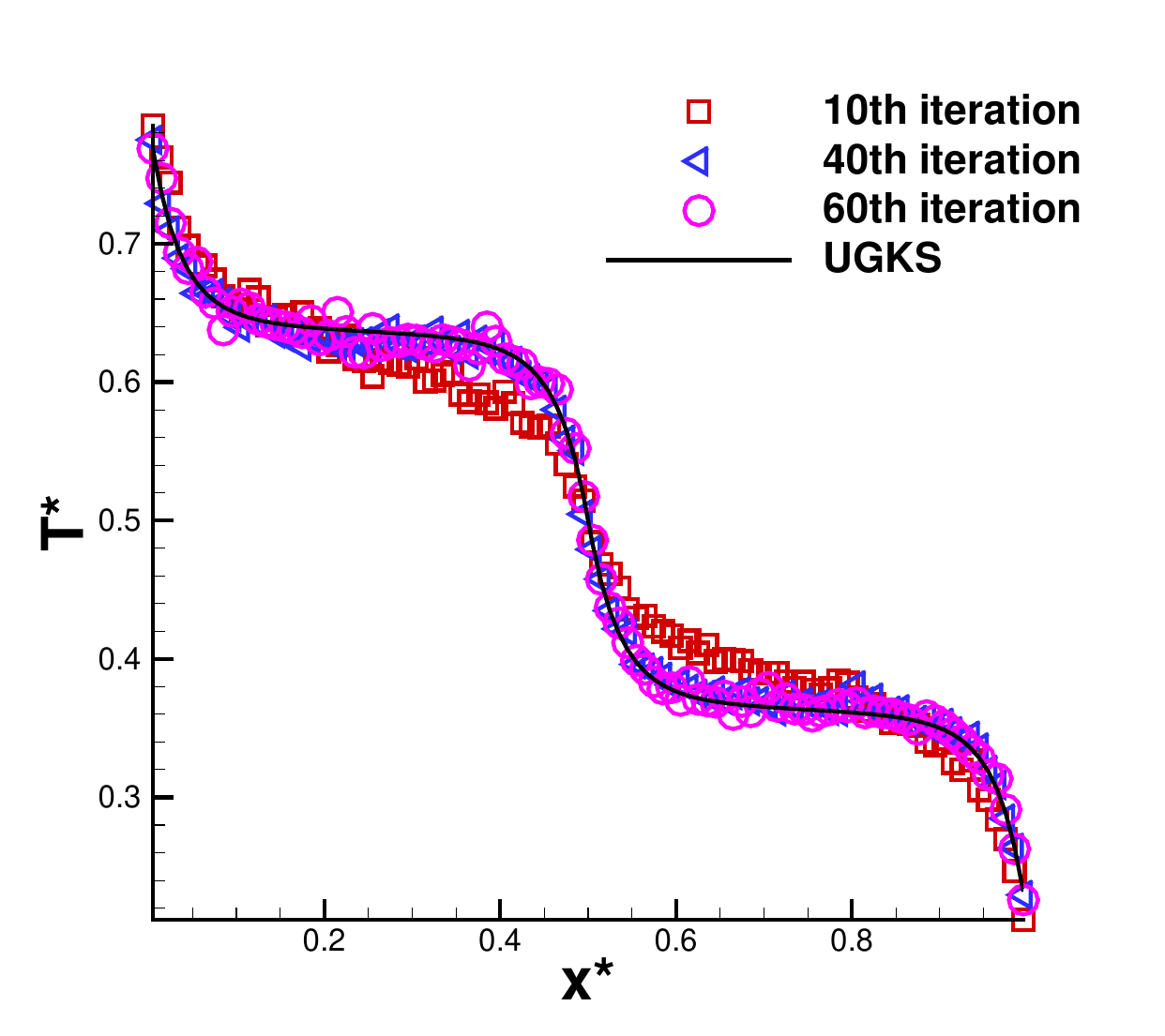}
	\caption{\label{multiscale-1d-heat-result}
		multi-scale case. }
\end{figure}

For the convergence rate, the UGKWP method requires 3800 steps to reach steady state, whereas the current method only needs 60 steps. Moreover, the current method's per-step speed is approximately four times faster than that of UGKWP. This translates to a 253-fold acceleration in CPU speed compared to the UGKWP method.

\subsection{Heat transfer in the 2D square domain}
To validate the effectiveness and fast convergence rate of the current method for phonon transport in multidimensional physical space, this section investigates the heat conduction problem in a 2D square domain across a range of Knudsen numbers.
Specifically, the top boundary is maintained at a high temperature $T_H$, while the remaining boundaries are held at a lower temperature $T_C$ as illustrated in
Fig~.\ref{2d-square}.

\begin{figure}[htp]	\label{2d-square}
	\centering	
	\includegraphics[height=0.45\textwidth]{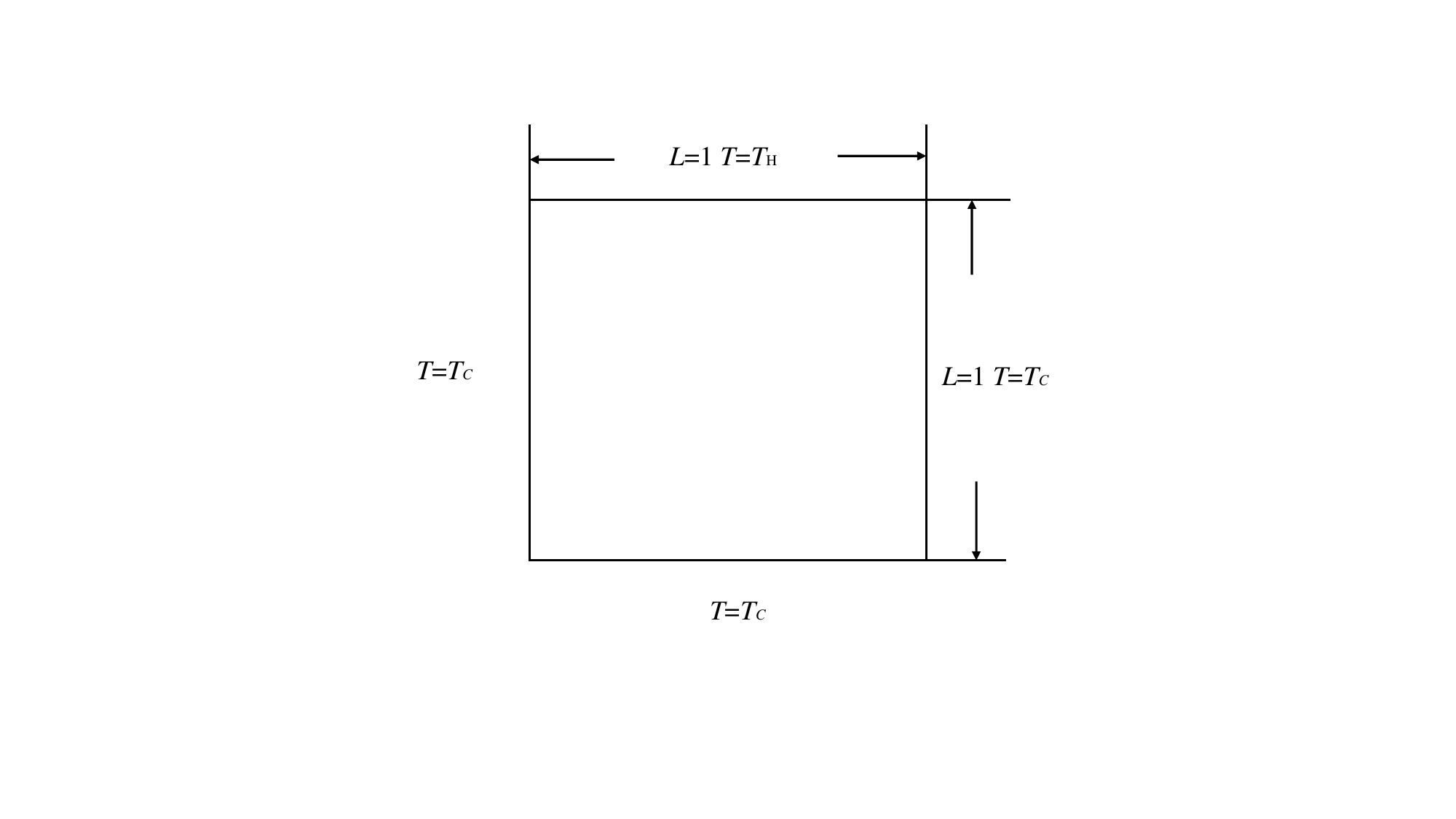}
	\caption{\label{2d-square}
		Computational domain and boundary conditions of Heat transfer in the 2D square domain. }
\end{figure}

In this case, we employ a two-dimensional uniform grid, discretized into 40 equally spaced points in each direction, resulting in a total of 1,600 grid points.
In this case, $N_{ref}=200$ in each cell for Kn = 10.0, 1.0, and $N_{ref}=100$ for Kn = 0.1, Kn = 0.01, balancing computational efficiency and statistical noise.
Furthermore, an additional 100 steps were incorporated for statistical averaging in the two-dimensional simulation to reduce statistical noise.
The computation results for different Knudsen numbers are shown in Fig~.\ref{2d-square-result}. The black solid lines in the figure represent the contour lines obtained using the current method, while the white dashed lines represent those computed with the DUGKS method. As can be seen, the current method exhibits very good agreement with the reference method in both the diffusive region and the ballistic region. From Table~\ref{square-table-without} and Table~\ref {square-table}, a two-order speedup is achieved for all cases except the statistical average costs. Overall, it achieves one order-of-magnitude acceleration.

\begin{figure}[htp]	\label{2d-square-result}
	\centering	
    \includegraphics[height=0.40\textwidth]{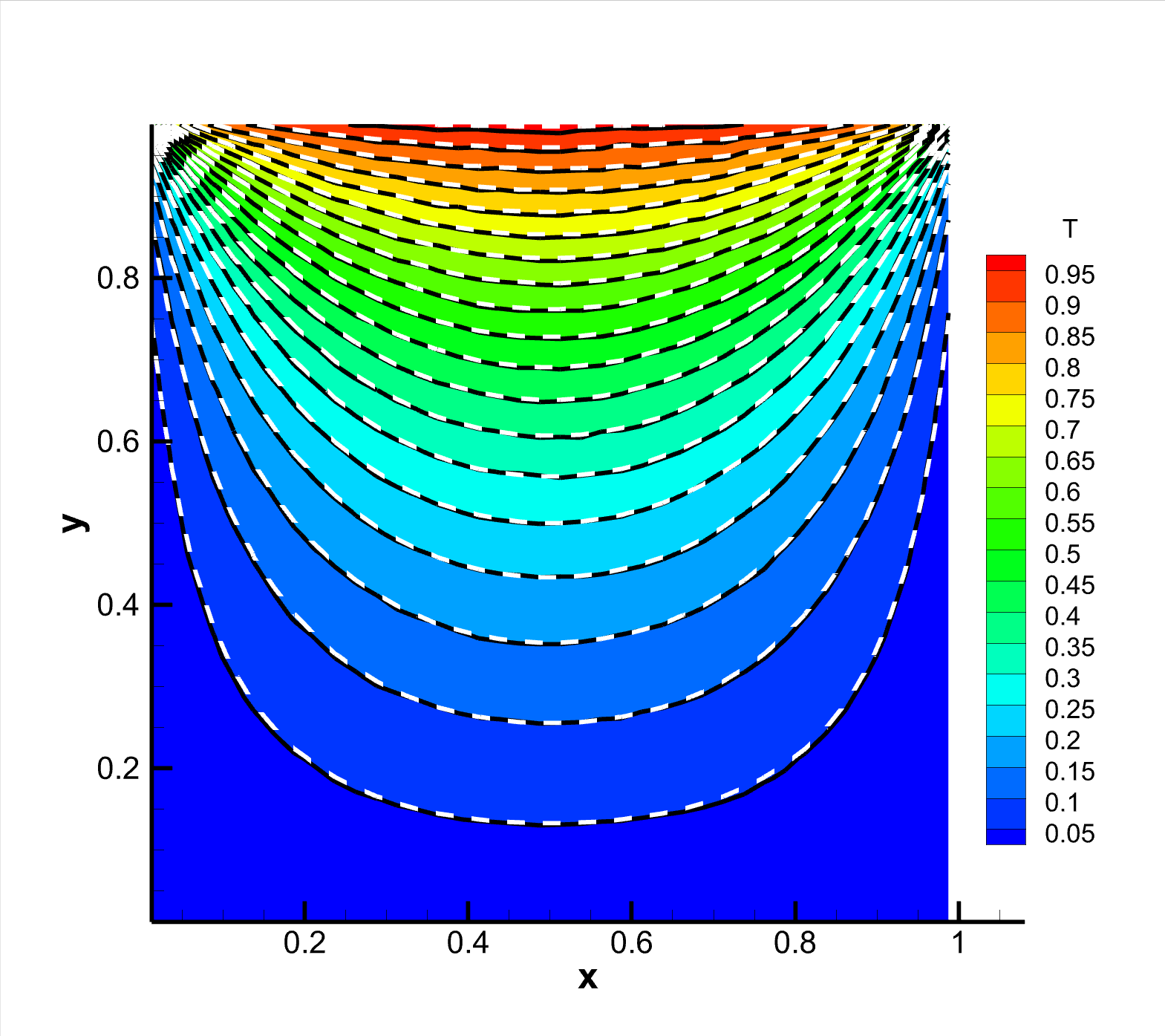}
	\includegraphics[height=0.40\textwidth]{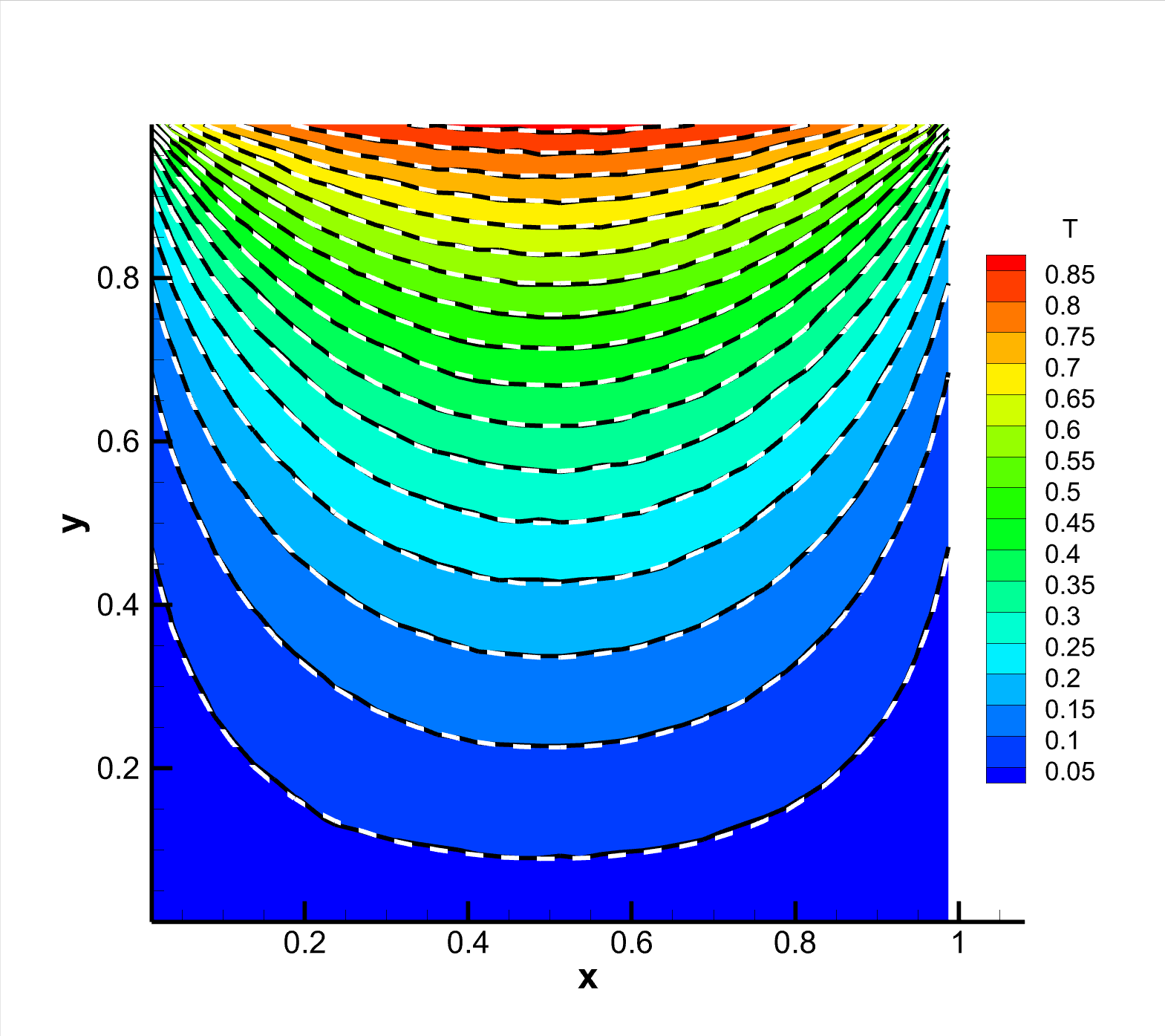}
    \includegraphics[height=0.40\textwidth]{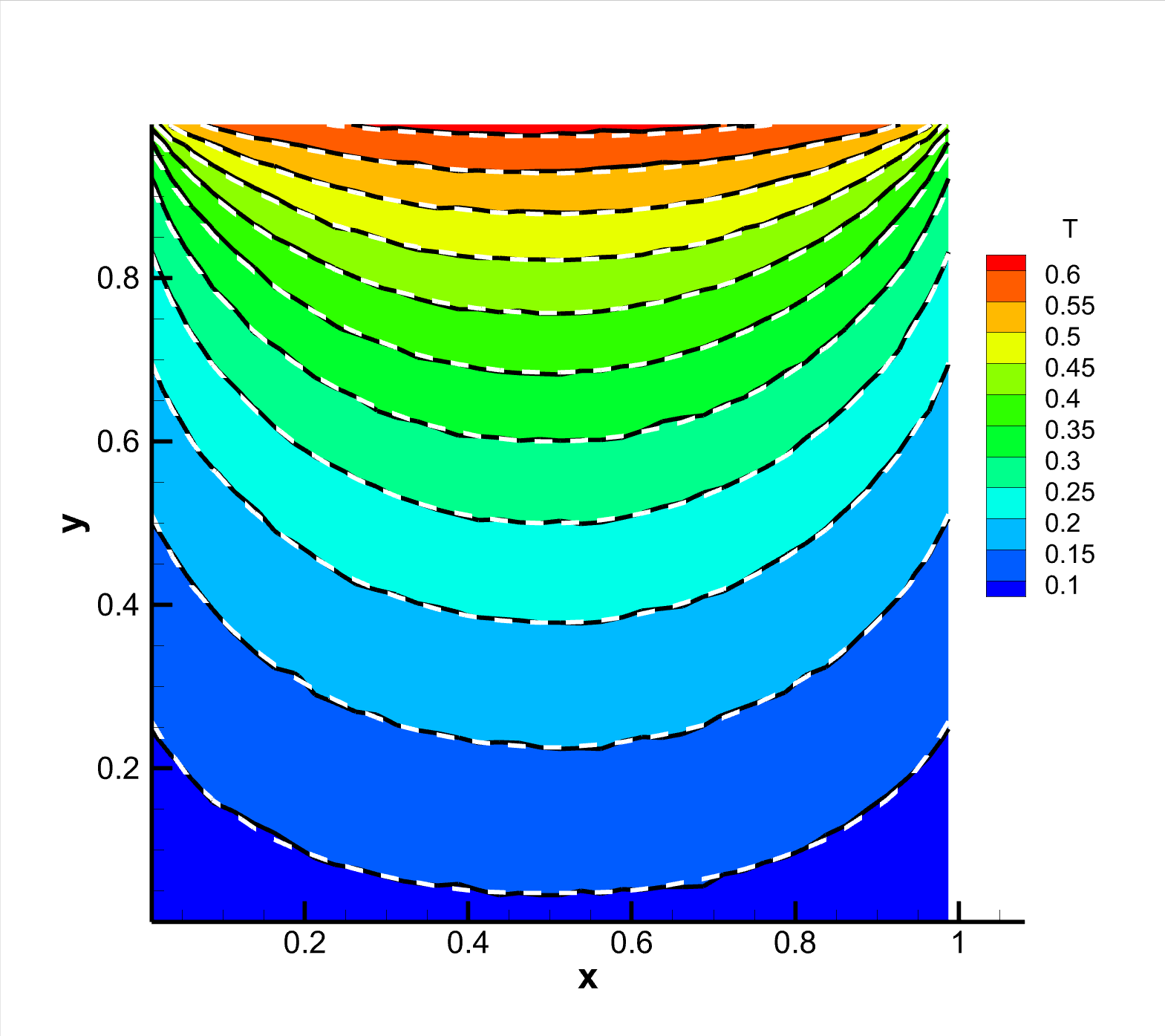}
    \includegraphics[height=0.40\textwidth]{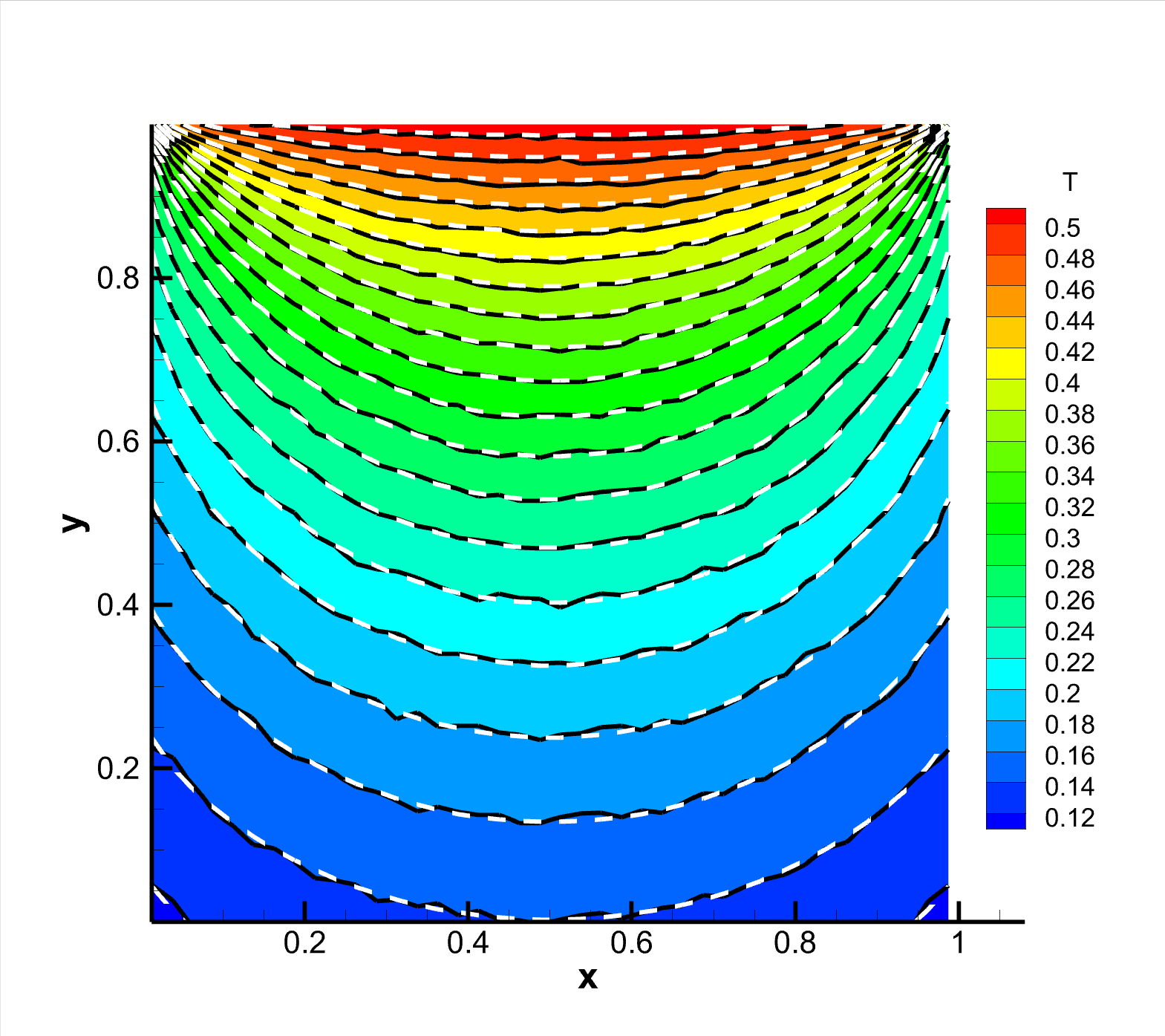}
	\caption{\label{2d-square-result}
		Comparison of 2D square heat transfer results, respectively corresponding: Kn = 0.01, Kn = 0.1, Kn = 1.0, Kn = 10.0. }
\end{figure}

\begin{table}[htp]
	\small
	\begin{center}
		\def\temptablewidth{1.0\textwidth}
		{\rule{\temptablewidth}{1pt}}
		\begin{tabular*}{\temptablewidth}{@{\extracolsep{\fill}}c|c|c|c|c|}
			Method & Kn = 10.0 &  Kn = 1.0 & Kn = 0.1 & Kn = 0.01  \\
			\hline
			UGKWP & 1200  & 1500  & 3600  &  12000   \\ 	
			Current & 10  & 10  & 40  &  400   \\ 	
                Speedup & 480  & 600  & 360  & 120   \\ 	
		\end{tabular*}
		{\rule{\temptablewidth}{0.1pt}}
	\end{center}
	\vspace{-4mm} \caption{\label{square-table-without} Comparison of the Computational costs of heat transfer in the 2D square domain between the UGKWP method and the current method (without statistical averaging).}
\end{table}

\begin{table}[htp]
	\small
	\begin{center}
		\def\temptablewidth{1.0\textwidth}
		{\rule{\temptablewidth}{1pt}}
		\begin{tabular*}{\temptablewidth}{@{\extracolsep{\fill}}c|c|c|c|c|}
			Method & Kn = 10.0 &  Kn = 1.0 & Kn = 0.1 & Kn = 0.01  \\
			\hline
			UGKWP & 1200 + 200 & 1500 + 200 & 3600 + 200 &  12000 + 200  \\ 	
			Current & 10 + 200 & 10 + 200 & 40 + 200 &  400 + 200  \\ 		
                Overall Speedup & 27 & 32 & 63 & 81  \\
		\end{tabular*}
		{\rule{\temptablewidth}{0.1pt}}
	\end{center}
	\vspace{-4mm} \caption{\label{square-table} Comparison of the Computational costs of Heat transfer in the 2D square domain between the UGKWP method and the current method (with 200 statistical averaging steps).}
\end{table}

\subsection{Heat transfer in the 2D rectangular domain}
This section investigates the heat transfer in a rectangular cavity with non-uniform temperature boundary conditions in the ballistic region. Computational domain and boundary conditions are shown in Fig~.\ref{2d-rec}
\begin{figure}[htb]	\label{2d-rec}
	\centering	
	\includegraphics[height=0.50\textwidth]{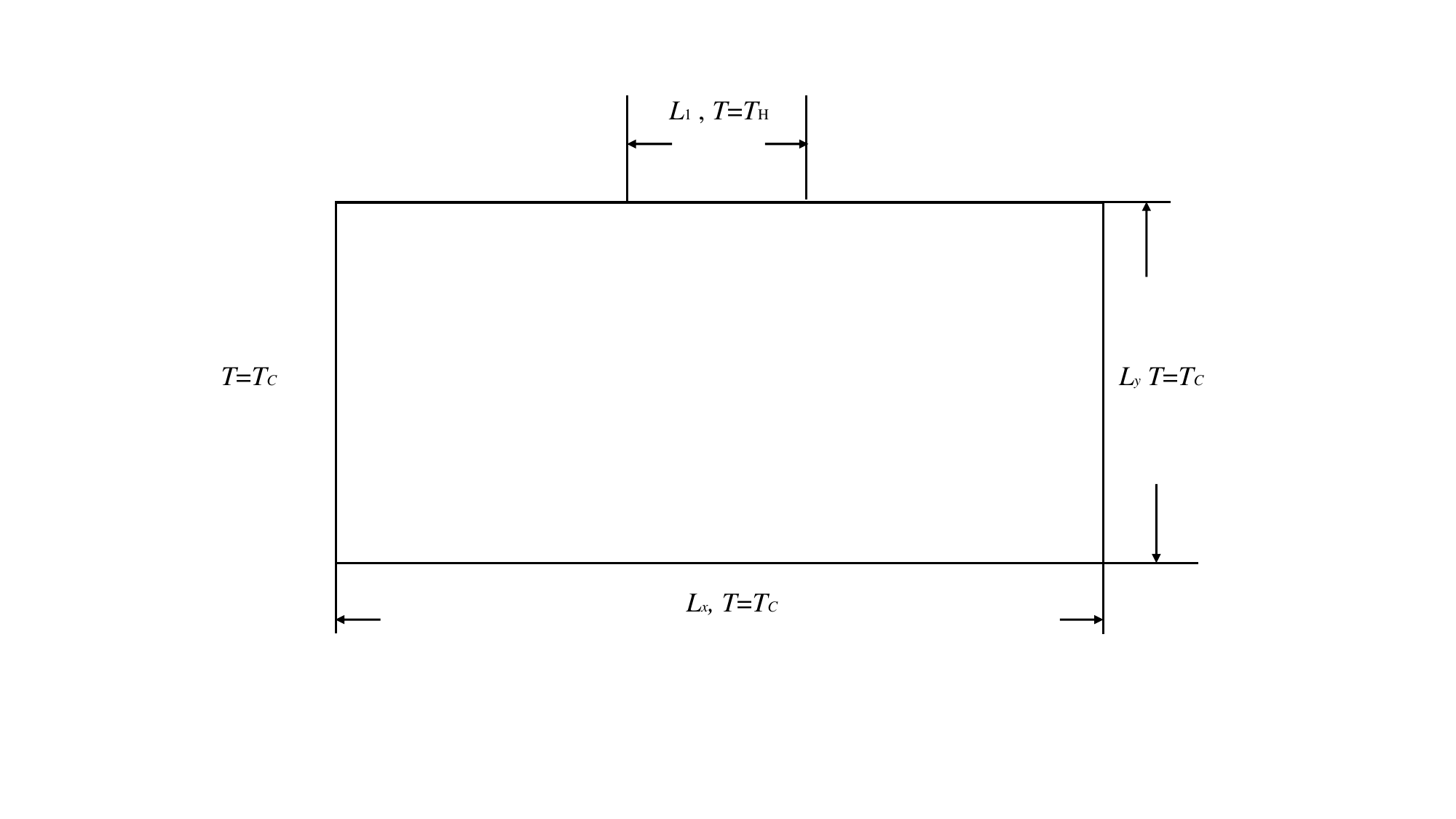}
	\caption{\label{2d-rec}
		Computational domain and boundary conditions of Heat transfer in the 2D rectangle domain. }
\end{figure}
where $L_x=2L_y=5L_1$. The heat source is positioned at the center of the upper boundary of the entire computational domain. It has a length of $L_1 =1$ and a temperature of $T_H$. Other boundaries' temperature is $T_C$.

In this computational domain, the x-direction is discretized into 100 grid cells. In comparison, the y-direction is discretized into 50 grid cells which means the whole domain is discretized into 5000 uniform grid cells.
Meanwhile, the reference particle sampling number for each cell is 200 to balance efficiency and statistical noise.
Moreover, like heat transfer in the 2D square domain, in this case, a 1500-statistic average step is also adopted to reduce statistic noise.

The computation results for $Kn =10.0$ and $Kn = 1.0$ are shown in Fig~.\ref{2d-square-result}. The black solid lines in the figure represent the contour lines obtained using the current method, while the white dashed lines represent those computed with the DUGKS method. As can be seen, the current method exhibits very good agreement with the reference method.
\begin{figure}[htp]	\label{2d-rec-result}
	\centering	
    \includegraphics[height=0.40\textwidth]{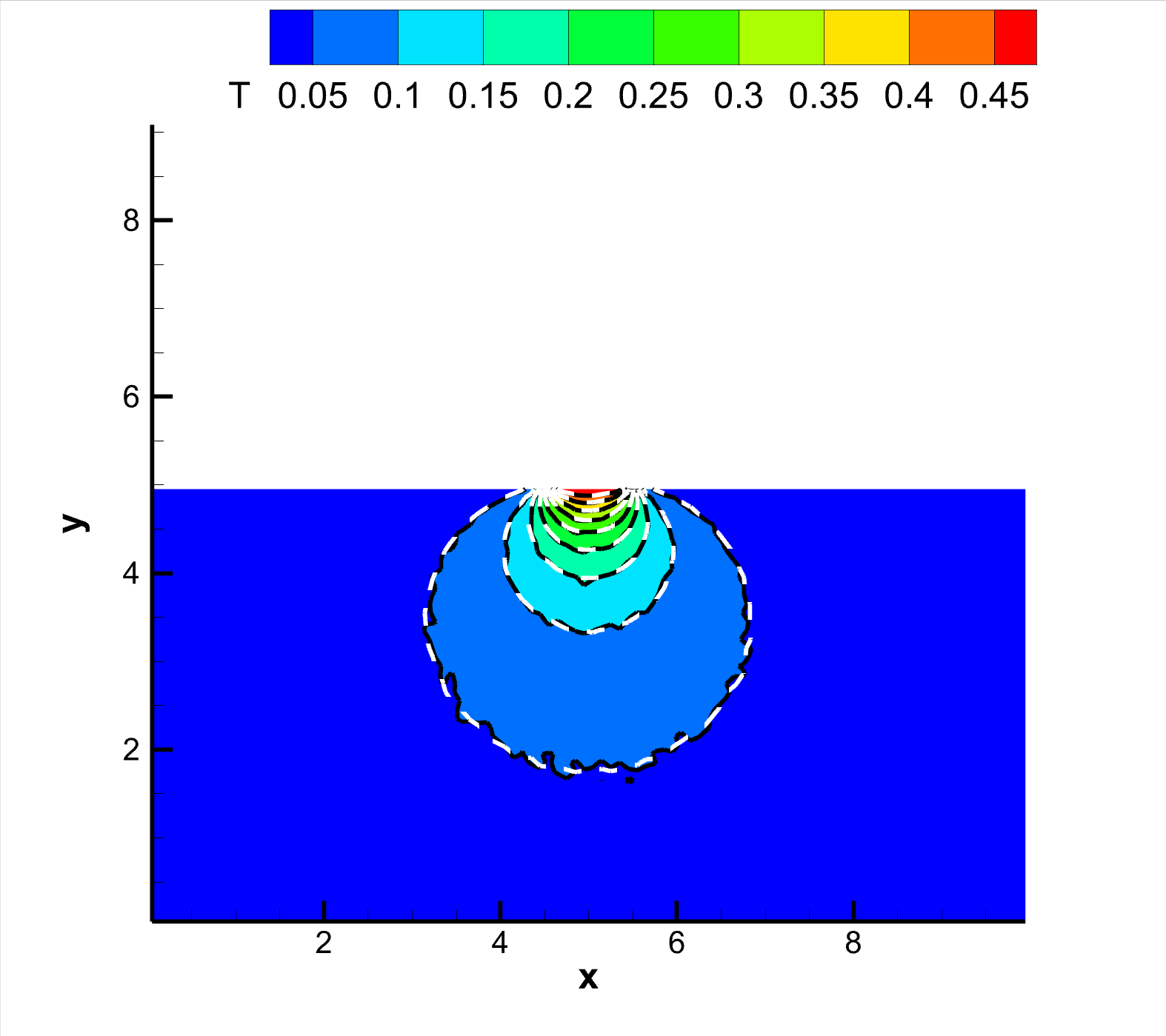}
	\includegraphics[height=0.40\textwidth]{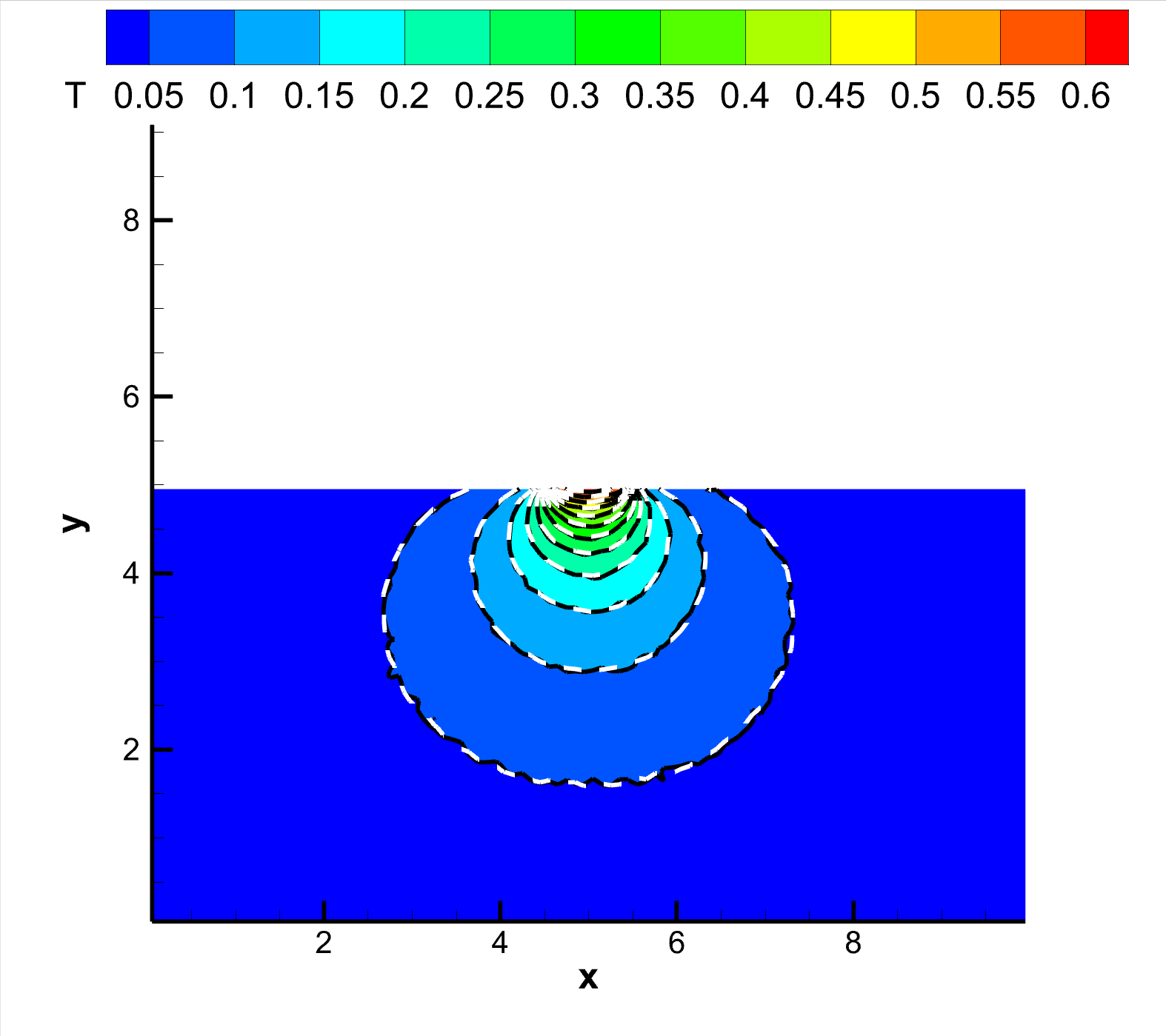}
	\caption{\label{2d-rec-result}
		Comparison of 2D rectangular heat transfer results. Left: Kn = 10.0, Right: 1.0. }
\end{figure}
From Table~\ref {rec-table} and Table~\ref{rec-table-without}, a two-order speedup is achieved for all cases except the statistical averaging costs. Overall, it achieves one order-of-magnitude acceleration.

\begin{table}[htp]
	\small
	\begin{center}
		\def\temptablewidth{1.0\textwidth}
		{\rule{\temptablewidth}{1pt}}
		\begin{tabular*}{\temptablewidth}{@{\extracolsep{\fill}}c|c|c|}
			Method & Kn = 10.0 &  Kn = 1.0   \\
			\hline
			UGKWP & 1500  & 1500    \\ 	
			Current & 10  & 10    \\ 	
                Speedup & 600  & 600    \\
		\end{tabular*}
		{\rule{\temptablewidth}{0.1pt}}
	\end{center}
	\vspace{-4mm} \caption{\label{rec-table-without} Comparison of the Computational costs of heat transfer in the 2D rectangular domain between the UGKWP method and the current method (without statistical averaging).}
\end{table}

\begin{table}[htp]
	\small
	\begin{center}
		\def\temptablewidth{1.0\textwidth}
		{\rule{\temptablewidth}{1pt}}
		\begin{tabular*}{\temptablewidth}{@{\extracolsep{\fill}}c|c|c|}
			Method & Kn = 10.0 &  Kn = 1.0   \\
			\hline
			UGKWP & 1500 + 200 & 1500 + 200   \\ 	
			Current & 10 + 200 & 10 + 200   \\ 	
                Overall Speedup & 32 & 32   \\
		\end{tabular*}
		{\rule{\temptablewidth}{0.1pt}}
	\end{center}
	\vspace{-4mm} \caption{\label{rec-table} Comparison of the Computational costs of Heat transfer in the 2D rectangular domain between the UGKWP method and the current method (with 200 statistical averaging steps).}
\end{table}

\section{Conclusion}

This paper introduces a novel steady-state particle acceleration method for solving multi-scale phonon transport. Unlike the previous explicit UGKWP method, our approach samples particles based on exact mean free paths at each step.
At large Knudsen numbers, particles travel longer distances, enabling faster convergence compared to the explicit UGKWP method. At small Knudsen numbers, the steady-state macroscopic prediction equation becomes dominant. The particle sampling from the equilibrium state significantly improves convergence speed in low Knudsen number regimes compared to the original explicit UGKWP method.
The paper also addresses the spatial-time-inconsistency problem by implementing a null collision method, which establishes a unified time scale when encountering large variations in particle mean free paths.
Numerical tests demonstrate the IUGKP method's excellent performance in accelerating multi-scale phonon transport solutions. The method achieves convergence speedups of one to two orders of magnitude, representing a significant advancement in multi-scale approaches for phonon transport problems.
Looking ahead, this IUGKP method shows promise for extension to photon transport, rarefied gas flow, and electron transport applications.

\section*{Acknowledgements}

The current research is supported by National Key R\&D Program of China (Grant Nos. 2022YFA1004500), National Science Foundation of China (12172316, 92371107), and Hong Kong research grant council (16301222, 16208324).

\bibliographystyle{unsrt}
\bibliography{hongyu}

\end{document}